\documentclass{article}

\usepackage{color,graphics,epsfig}

\newcommand{\graph}[6]{
 \begin{center}
  \vspace{#1}
  \protect\hspace*{#2} \mbox{\resizebox{#4}{#5}{\includegraphics{#6}}}
  \vspace{#3}
 \end{center} 
                      }

\def\greaterthansquiggle{\raise.3ex\hbox{$>$\kern-.75em\lower1ex\hbox{$\sim$}}}
\def\lessthansquiggle{\raise.3ex\hbox{$<$\kern-.75em\lower1ex\hbox{$\sim$}}}
\newcommand{\beq}{\begin{equation}}
\newcommand{\eeq}{\end{equation}}
\newcommand{\beqa}{\begin{eqnarray}}
\newcommand{\eeqa}{\end{eqnarray}}
\newcommand{\ba}{\begin{array}}
\newcommand{\ea}{\end{array}}
\newcommand{\no}{\nonumber}

\newcommand{\ra}{\rightarrow}

\newcommand{\cL}{{\cal L}}
\newcommand{\M}{{\cal M}}

\newcommand{\cP}{{\cal P}}

\def\noi {\noindent}

\def\a               {\alpha}
\def\b               {\beta}

\def\d               {\delta}

\def\s               {\sigma}
\def\t               {\theta}
\def\x               {\chi}

\def\D               {\Delta}

\def\ti    {\tilde}
\def\sf    {{\ti f}}
\def\sq    {{\ti q}}
\def\st    {{\ti t}}
\def\sb    {{\ti b}}
\def\stau  {{\ti\tau}}
\def\snu   {{\ti\nu}}

\def\ch    {\ti \x}
\def\nt    {\ti \x^0}
\def\sg    {\ti g}

\def\cth   {\cos\theta}

\def\tsf   {\theta_{\ti f}}
\def\tst   {\theta_{\ti t}}
\def\tsb   {\theta_{\ti b}}

\def\cst   {\cos\theta_{\ti t}}
\def\csb   {\cos\theta_{\ti b}}
\def\cstau {\cos\theta_{\ti\tau}}

\newcommand{\msf}[1]   {m_{\ti f_{#1} }}
\newcommand{\mst}[1]   {m_{\ti t_{#1} }}
\newcommand{\msb}[1]   {m_{\ti b_{#1} }}
\newcommand{\mstau}[1] {m_{\ti \tau_{#1} }}
\newcommand{\mnt}[1]   {m_{\ti \chi^0_{#1} }}
\newcommand{\mch}[1]   {m_{\ti \chi^+_{#1} }}

\newcommand{\msg}      {m_{\ti g}}

\def\Pm  {{\cal P}_-^{}}
\def\Pp  {{\cal P}_+^{}}
\def\fbi {{\rm fb}^{-1}}

\newcommand{\gsim}{\;\raisebox{-0.9ex}
           {$\textstyle\stackrel{\textstyle >}{\sim}$}\;}

\newcommand{\lsim}{\;\raisebox{-0.9ex}{$\textstyle\stackrel{\textstyle<}
           {\sim}$}\;}

\newcommand{\recht} {\begin{picture}(4,4)
                      \linethickness{1.7mm}
                      \put(1,0.25){\line(0,1){2}}
                      \thinlines
                     \end{picture}
                    }

\newcommand{\rechtl} {\begin{picture}(4,4)
                      \put(1,0.25){\framebox(2,2){}}
                     \end{picture}
                    }

\begin{document}  
\setlength{\unitlength}{1mm}

\begin{flushright}
  UWThPh-2000-06\\
  HEPHY-PUB 729/00\\
  FTUV/00-10\\ 
  IFIC/00-10\\
  hep-ph/0002115\\
  LC-TH-2000-031\\[3mm]
\end{flushright}

\begin{center}

{\Large \bf\boldmath Phenomenology of \\[1mm]
        Stops, Sbottoms, $\tau$-Sneutrinos, and Staus \\[2mm]
        at an $e^+e^-$ Linear Collider} 

\vspace{5mm}

{\large A.~Bartl,$^1$~ 
H.~Eberl,$^2$~ 
S.~Kraml,$^2$~ 
W.~Majerotto,$^2$~
W.~Porod\,$^3$} \\

\vspace{4mm}

{\normalsize \it
$^1$~Institut f\"ur Theoretische Physik, Universit\"at Wien, 
     A--1090 Vienna, Austria \\
$^2$~Inst. f. Hochenergiephysik,  
     \"Osterr. Akademie d. Wissenschaften, 
     A--1050 Vienna, Austria \\
$^3$~Inst.~de F\'\i sica Corpuscular (IFIC), CSIC, 
     E--46071 Val\`encia, Spain} 

\end{center}

\begin{abstract} 
We discuss production and decays of stops, sbottoms,
$\tau$-sneutrinos, and staus in $e^+e^-$ annihilation 
in the energy range $\sqrt{s} = 0.5-1$~TeV. We present numerical
predictions within the Minimal Supersymmetric Standard
Model for cross sections and decay rates, including one--loop 
radiative corrections as well as initial state
radiation. We also study the importance of beam polarization for the
determination of the underlying SUSY parameters. 
Moreover, we make a comparison of the potential to study squarks and
sleptons of the 3rd generation between Tevatron, LHC, and Linear Collider. 
\end{abstract}
\renewcommand{\thefootnote}{\fnsymbol{footnote}}
\footnotetext{Contribution to the Proceedings of the 
``2nd Joint ECFA/DESY Study on Physics and Detectors for a Linear 
  Electron--Positron Collider''.}

\renewcommand{\thefootnote}{\arabic{footnote}}

\section{Introduction}
In supersymmetric (SUSY) extensions of the Standard Model (SM)
squarks $\ti q_L^{}$, $\ti q_R^{}$, sleptons $\ti \ell_L$, $\ti
\ell_R$, and sneutrinos $\ti \nu_{\ell}$ are introduced as the
scalar partners of the quarks $q_{L,R}^{}$, leptons $\ell_{L,R}$, and 
neutrinos $\nu_{\ell}$ \cite{nilles}. For each sfermion of
definite flavour 
the states $\ti f_L$ and $\ti f_R$ are interaction states which
are mixed by 
Yukawa terms. The mass eigenstates are denoted by $\ti f_1$ and
$\ti f_2$ (with the convention $m_{\ti f_1} < m_{\ti f_2}$).
Strong $\ti f_L - \ti f_R$ mixing is expected for the third
generation sfermions, because in this case 
the Yukawa couplings can be large. In particular, in the sector
of the scalar top quarks these mixing effects will be large due
to the large top quark mass. The lighter mass eigenstate 
$\ti t_1$ will presumably be the lightest squark state
\cite{elru,stop}. If the 
SUSY parameter $\tan \beta$ is large, $\tan \beta \gsim 10$,
then also $\ti b_L - \ti b_R$ and $\ti \tau_L - \ti \tau_R$
mixing has to be taken into account and will lead to observable
effects \cite{sbot,bamapo}. The 
experimental search for the third generation sfermions is
an important issue at all present and future colliders. It will
be particularly interesting at an $e^+e^-$ Linear Collider with
center of mass energy $\sqrt{s} = 0.5-1.5$~TeV, where these
states are expected to be pair produced. Moreover, at an
$e^+e^-$ Linear Collider with this energy and an integrated
luminosity of about $500$~fb$^{-1}$ it will be possible
to measure masses, cross sections and decay branching
ratios with high precision \cite{acco}. 
This will allow us to obtain information on 
the fundamental soft SUSY breaking parameters. 
Therefore, it is necessary to investigate how this
information can 
be extracted from the experimental data, and how precisely these
parameters can be determined. In this way it will be possible to
test our theoretical ideas about the underlying SUSY breaking
mechanism. 

Phenomenological studies on SUSY particle searches at the LHC have shown 
that the detection of the scalar top quark may be very difficult due 
to the overwhelming background from $t\bar{t}$ production 
\cite{dydak,gianotti,polesello,cms,atlas}. 
This is in particular true for $\mst{1} \lsim 250$~GeV \cite{dydak}.  
In principle, such a light stop could be discovered at the Tevatron. 
The actual mass reach, however, strongly depends on the luminosity,  
decay modes, and the available phase--space \cite{teva,demina}.
Thus an $e^+e^-$ Linear Collider with $\sqrt{s} \sim 500$~GeV could 
even be a discovery machine for $\st_1$. 

In this contribution we summarize the phenomenology of 
$\st$, $\sb$, $\stau$, and $\snu_\tau$ in 
$e^+e^-$ annihilation at energies between $\sqrt{s} = 500$~GeV
and $1$~TeV. We give numerical results for the production
cross sections taking into account polarization of both the $e^-$
and $e^+$ beams. In particular, we show that by using polarized
beams it will be possible to determine the fundamental SUSY
parameters with higher precision than without polarization. 
Moreover, we discuss the decays of these particles.
The production cross sections as well as the 
decay rates of the sfermions show a distinct dependence on the
$\ti f_L$--$\ti f_R$ mixing angles. 
Squarks (sleptons) can decay into quarks (leptons) plus 
neutralinos or charginos. Squarks may also decay into gluinos. 
In addition, if the splitting between the different sfermion 
mass eigenstates is large enough, transitions between these states 
by emmission of weak vector bosons or Higgs bosons are possible. 
These decay modes can be important for the higher mass
eigenstates, and lead to complicated cascade decays. 
In the case of the lighter stop, however, all these tree--level 
two--body decays may be kinematically forbidden. 
Then the $\ti t_1$ has more complicated higher--order 
decays \cite{hikasa-kobayashi,werner3bdy,djouadi4bdy}. 

The framework of our calculation is the Minimal Supersymmetric
Standard Model (MSSM) \cite{nilles} which contains the Standard
Model (SM) particles 
plus the sleptons $\ti \ell^{\pm}$, sneutrinos $\ti \nu_{\ell}$, 
squarks $\ti q$, gluinos $\ti g$, two pairs of charginos 
$\ti \chi^{\pm}_i$, $i = 1, 2$, four neutralinos, $\ti \chi^0_i$, 
$i = 1,\ldots,4$, and five Higgs bosons, $h^0$, $H^0$, $A^0$,
$H^\pm$ \cite{guha}.

In Section~2 we shortly review the basic features of left--right
mixing of squarks and sleptons of the 3rd generation,   
and present formulae and numerical results 
for the production cross sections with polarized $e^-$ and $e^+$
beams. 
In Section~3 we discuss the decays of these particles and present 
numerical results for their branching ratios. 
In Section~4 we give an estimate of
the errors to be expected for the fundamental soft
SUSY--breaking parameters of the stop mixing matrix. 
In Section~5 we compare the situation concerning
stop, sbottom, and stau searches at LHC and Tevatron with that at an
$e^+e^-$ Linear Collider. 
Section~6 contains a short summary.

\section{Production Cross Sections}


Left--right mixing of the sfermions is described by the 
symmetric $2 \times 2$ mass matrices which in the 
$(\ti f_L, \ti f_R)$ basis $(f = t, b, \tau )$ 
read \cite{elru,guha}
\beq
\M^2_{\ti f} = \left(
\ba{cc}
M^2_{\ti f_L} & a_f m_f \\
a_f m_f & M^2_{\ti f_R} 
\ea 
\right) \,.
\label{xyz}
\eeq
The diagonal elements of the sfermion mass matrices are 
\beqa
M^2_{\ti f_L} &=& M^2_{\ti F} + 
m_Z^2 \cos 2\beta(T^3_f - e_f \sin^2\Theta_W) + m^2_f \,, \label{ML}\\
M^2_{\ti f_R} &=& M^2_{\ti F'} + 
 e_f m_Z^2 \cos 2\beta \sin^2\theta_W + m^2_f \label{MR}
\eeqa
where $m_f$, $e_f$ and $T^3_f$ are the mass, charge and third
component of weak isospin of the fermion $f$, and $\theta_W$ is
the Weinberg angle. Moreover, $M_{\ti F}=M_{\ti Q}$ for 
$\ti f_L = \ti t_L, \ti b_L$, $M_{\ti F}=M_{\ti L}$ for
$\ti f_L = {\ti \tau}_L$, and $M_{\ti F'} = M_{\ti U}, M_{\ti D},
M_{\ti E}$ for $\ti f_R = \ti t_R, \ti b_R, \ti \tau_R$,
respectively. $M_{\ti Q}$, $M_{\ti U}$, $M_{\ti D}$, $M_{\ti
L}$, and $M_{\ti E}$ are soft SUSY--breaking mass parameters
of the third generation sfermion system.
The off--diagonal elements of the sfermion mass matrices are 
\beqa
  m_t a_t &=& m_t(A_t - \mu \cot \beta)\,, \label{At}\\ 
  m_b a_b &=& m_b(A_b - \mu \tan \beta)\,, \label{Ab}\\
  m_{\tau} a_{\tau} &=& m_{\tau}(A_{\tau} - \mu \tan\beta)
\label{Atau} 
\eeqa 
for stop, sbottom, and stau, respectively. $A_t$, $A_b$,
$A_{\tau}$ are soft SUSY--breaking trilinear scalar coupling
parameters. Evidently, in the stop sector there can
be strong $\ti t_L$-$\ti t_R$ mixing due to the large top quark 
mass. In the case of sbottoms and staus the $\ti f_L - \ti f_R$
mixing effects are also non-negligible if
$\tan \beta \gsim 10$. We assume that all parameters are
real. Then the mass matrices can be diagonalized by $2 \times 2$
orthogonal matrices. The mass eigenvalues for the sfermions
$\ti f = \ti t, \ti b, \ti \tau$  are
\beq
  m^2_{\ti f_{1,2}} = \frac{1}{2} (M^2_{\ti f_L} + M^2_{\ti f_R}
  \mp \sqrt{(M^2_{\ti f_L} - M^2_{\ti f_R})^2 + 4m_f^2 a^2_f}\,) \,,
\label{sfmass}
\eeq
and the mass eigenstates are
\beqa
  \ti f_1^{} &=& \ti f_L^{} \cth_{\ti f} + \ti f_R^{}\sin\theta_{\ti f}\,, \\
  \ti f_2^{} &=& \ti f_R^{} \cth_{\ti f} - \ti f_L^{}\sin\theta_{\ti f}\,,
\eeqa
where $\ti t_1$,  $\ti b_1$, $\ti \tau_1$ denote the lighter eigenstates.
The sfermion mixing angle is given by
\beq
  \cos \theta_{\ti f} = \frac{-a_f m_f}
    {\sqrt{(M^2_{\ti f_L} - m^2_{\ti f_1})^2 + a_f^2 m_f^2}}\,, \qquad 
  \sin \theta_{\ti f} = \frac{M^2_{\ti f_L} - m^2_{\ti f_1}}
    {\sqrt{(M^2_{\ti f_L} - m^2_{\ti f_1})^2 + a_f^2 m_f^2}}\,.
\label{eq:mixangle}
\eeq
The $\snu_\tau$ appears only in the left--state. 
Its mass is  
\beq
  m^2_{\ti\nu_\tau} = M^2_{\ti L} + \frac{1}{2} m^2_Z\cos 2\beta \,.
\eeq


The reaction $e^+ e^- \ra \ti f_i \bar{\ti f_j}$ proceeds
via $\gamma$ and $Z$ exchange in the s--channel.
For polarized $e^-$ and $e^+$ beams the cross section of this
reaction at tree level has the form \cite{sabine}
\beqa
  \sigma^0 &=& \frac{\pi \alpha^2 \kappa^3_{ij}}{s^4}
  \left\{ e^2_f \delta_{ij}(1-\cP_-\cP_+) -
  \frac{e_f c_{ij} \delta_{ij}}{2 s^2_W c^2_W}
  \left[\, v_e(1-\cP_-\cP_+) - a_e(\cP_- - \cP_+)\,\right]\,D_{\gamma Z}\right. 
  \no \\
  && \left. \mbox{} \hspace{24mm} + \frac{c^2_{ij}}{16 s^4_W c^4_W}
  \left[\,(v^2_e + a^2_e)(1-\cP_-\cP_+) - 2 v_e a_e(\cP_- - \cP_+)\,\right]\,
  D_{ZZ} \right\},
\label{eq:sig0}
\eeqa
where $\cP_-$ and $\cP_+$ denote the degree of polarization 
of the $e^-$ and $e^+$ beams, with the convention $\cP_{\pm}=-1,0,+1$
for left--polarized, unpolarized, right--polarized $e^{\pm}$
beams, respectively. 
(E.g., $\cP_-=-0.9$ means that 90\% of the electrons are 
left--polarized and the rest is unpolarized.)
$v_e = 4s^2_W - 1$, $a_e = -1$ are the vector
and axial--vector couplings of the electron to the $Z$, 
$s^2_W \equiv \sin^2 \theta_W$, $c^2_W \equiv \cos^2 \theta_W$, and
$c_{ij}$ is the $Z\ti f_i \ti f_j$ coupling (up to a factor $1/\cos\t_W$)
\beq
  c_{ij}^{}=\left( \begin{array}{cc} 
  T^3_f\cos^2 \theta_{\sf}-e_f s^2_W & -\frac{1}{2}T^3_f\sin 2\theta_{\sf}\\ 
  -\frac{1}{2}T^3_f\sin 2\theta_{\sf} & T^3_f\sin^2 \theta_{\sf}-e_f s^2_W
  \end{array}\right)\,. 
\label{eq:cij}
\eeq
Furthermore, in Eq.~(\ref{eq:sig0}) $\sqrt{s}$ is the
center--of--mass energy, 
$\kappa_{ij}=[(s-\msf{i}^2-\msf{j}^2)^2 - 4\msf{i}^2\msf{j}^2]^{1/2}$, and
\beq
  D_{ZZ} = \frac{s^2}{(s-m^2_Z)^2 + \Gamma^2_Z m^2_Z}\,, \qquad
  D_{\gamma Z} = \frac{s(s-m^2_Z)}{(s-m^2_Z)^2 + \Gamma^2_Z m^2_Z}\,.
\eeq

\noi
The cross section in Eq. (\ref{eq:sig0}) depends
on the sfermion mixing parameters, because the 
$Z\,\ti f_i \bar{\ti f_j}$ couplings Eq.~(\ref{eq:cij})) 
depend on the mixing angles. For example, the couplings  
$Z\,\ti t_1 \bar{\ti t_1}$, $Z\ti b_1 \bar{\ti b_1}$, and 
$Z\,\ti \tau_1 \bar{\ti \tau_1}$ vanish at 
$\theta_{\ti t} = 0.98$, $\theta_{\ti b} = 1.17$, and 
$\theta_{\ti \tau} = 0.82$, respectively. 
There is a destructive interference between the $\gamma $ and
$Z$--exchange contributions that leads to
characteristic minima of the cross sections 
at specific values of the mixing angles $\tsf$, 
which according to Eq. (\ref{eq:sig0}) depend on $\sqrt{s}$ and on
the beam polarizations $\cP_-$ and $\cP_+$ \cite{lcstop}.

In Figs.~\ref{fig:Xsectsq}\,a,\,b we show the $\sqrt{s}$ 
dependence of the stop and sbottom pair production cross sections for 
$\mst{1}=220$ GeV, $\mst{2}=450$ GeV, $\cst=-0.66$, 
$\msb{1}=284$ GeV, $\msb{2}=345$ GeV, $\csb=0.84$, 
and $\msg=555$ GeV. Here we have included supersymmetric 
QCD (i.e. gluon and gluino) corrections \cite{glucorr,helmut} and initial 
state radiation (ISR) \cite{isr}. 
\footnote{The Fortran program \cite{calv} is available on the Web.}
The latter typically changes the cross section by $\sim 15\%$.
The relative importance of the gluon and gluino corrections can be 
seen in Figs.~\ref{fig:Xsectsq}\,c,\,d where 
we plot $\D\s/\s^0$ for $\ti t_1\bar{\ti t_2}$ and $\ti b_1\bar{\ti b_1}$ 
production for the parameters of Fig.~\ref{fig:Xsectsq}\,a,\,b. 
In addition we also show the leading electroweak corrections in order 
of Yukawa couplings squared \cite{yuk} 
for $M=200$~GeV, $\mu=800$~GeV, $m_A=300$~GeV, and $\tan\b=4$. 
Let us discuss these corrections in more detail:


\begin{figure}[p]
\graph{0cm}{0cm}{-0.6cm}{!}{!}{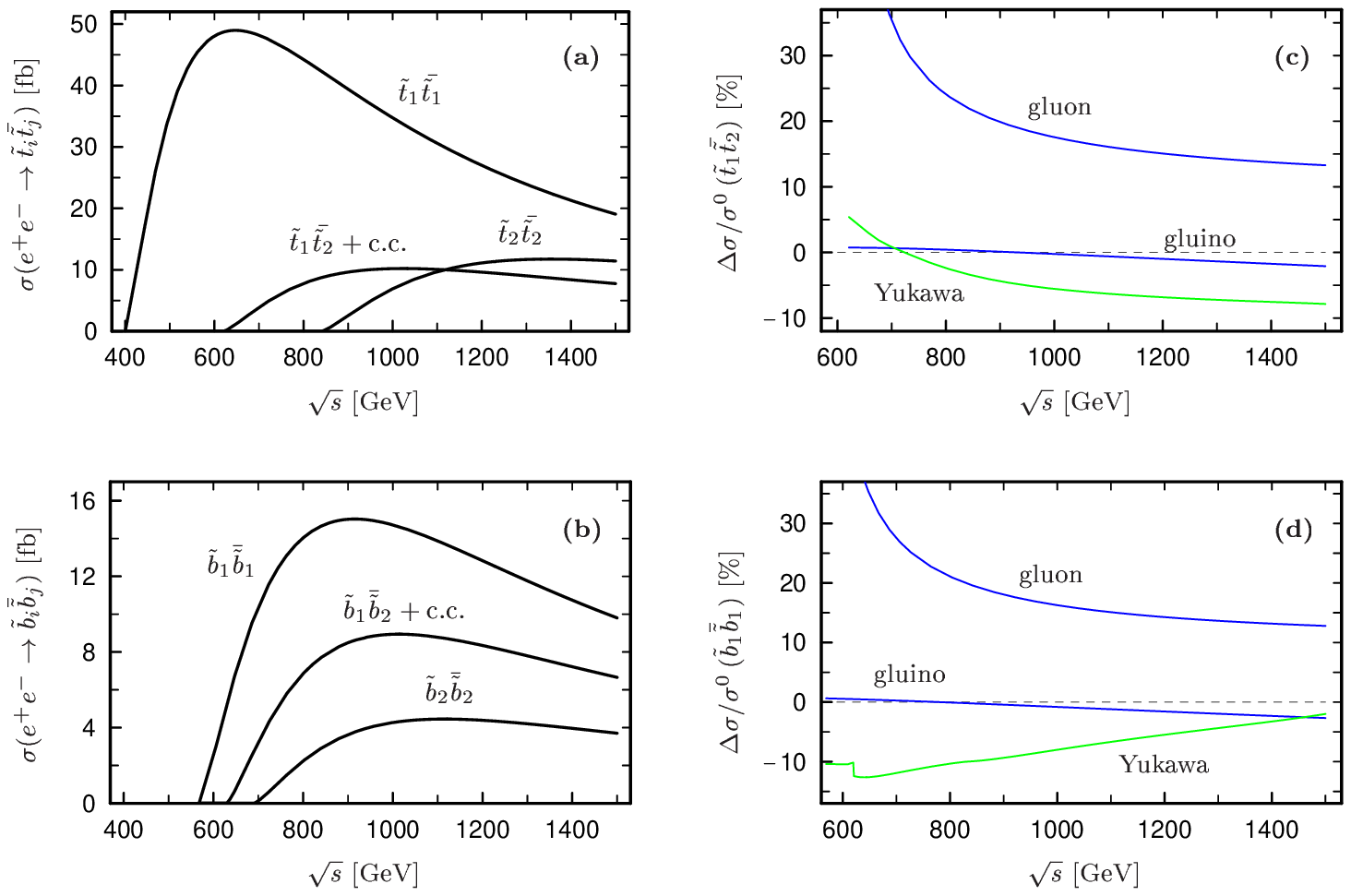}
\caption{{\bf (a,\,b)} Total cross sections for 
  $e^+e^-\to\st_i\bar\st_j$, $\sb_i\bar\sb_j$ 
  as a function of $\sqrt{s}$ for 
  $\mst{1}=200$ GeV, $\mst{2}=420$ GeV, $\cst=-0.66$, 
  $\msb{1}=284$ GeV, $\msb{2}=345$ GeV, $\csb=0.84$, and 
  $\msg=555$ GeV; included are SUSY--QCD and ISR corrections. 
  {\bf (c,\,d)} gluon, gluino, and Yukawa coupling corrections 
  relative to the tree level cross section of $\st_1\bar{\st_2}$ 
  and $\sb_1\bar\sb_1$ production  
  for $M=200$ GeV, $\mu=800$ GeV, $\tan\b=4$, $m_A=300$ GeV, 
  and the other parameters as in (a,\,b).}
\label{fig:Xsectsq}
\end{figure}

\begin{figure}[p]
\graph{0cm}{0cm}{-0.7cm}{!}{!}{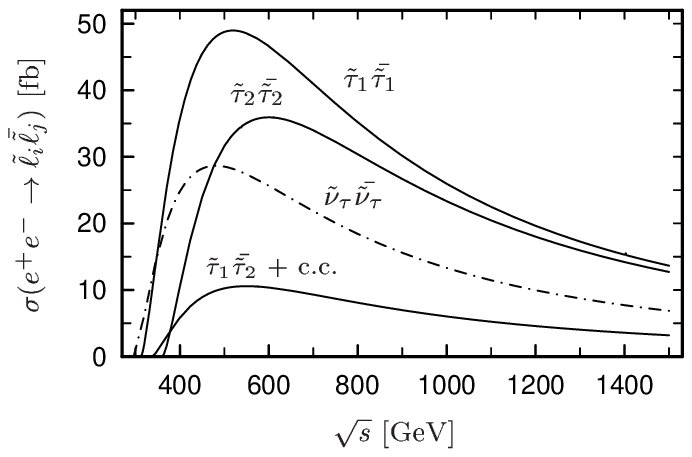}
\caption{Total cross sections of stau and sneutrino pair production
  as a function of $\sqrt{s}$ for 
  $\mstau{1}=156$ GeV, $\mstau{2}=180$ GeV, $\cstau=0.77$, and 
  $m_{\snu_\tau}=148$ GeV.}
\label{fig:Xsectsl}
\end{figure}

\noi
The standard QCD correction \cite{glucorr} (due to virtual gluon 
exchange and real gluon emission) is proportional to the tree--level cross 
section: $\s = \s^0\,(1 + \frac{4\a_s}{3\pi}\D)$ with $\D$ 
depending on the velocity of the outgoing squarks. 
In the high energy limit $\beta= 1-4m_{\sq_i}^2/s \to 1$ we have $\D=3$, 
i.e. the gluonic correction amounts to 10--15\% of $\s^0$.  
Notice that this is four times the corresponding correction for 
quark production. 
At/near the threshold, colour--Coulomb effects have to be taken 
into account \cite{bigi-fadin-khoze}. 
These lead to $\D\simeq\pi^2/(2\b)-2$ near the threshold.
Very close to threshold the perturbation expansion becomes unreliable, 
and the non--perturbative contribution leads to a constant cross 
section for $\b=0$. Moreover, bound state formation is felt in this region. 
A recent study \cite{antonelli} concluded that these bound states cannot 
be detected at an $e^+e^-$ Linear Collider. 
Still they may affect the precision of a mass determination of squarks 
by threshold scans. 
(Further investigations are necessary for quantitative results.) 
On the other hand, measuring the $\b^3$ rise of the cross section, 
as well as the $\sin^2\vartheta$ dependence of the differential cross section 
($\vartheta$ being the scattering angle), 
will be useful for confirming the spin-0 character of squarks and sleptons.  

\noi
The gluon correction has clearly the largest effect. 
However, for precision measurements also gluino exchange \cite{helmut}  
has to be taken into account. In contrast to the former, which is 
always positive, the gluino correction can be of either sign. 
Moreover, it does not factorize with the tree level but leads to an 
additional dependence on the squark mixing angle.
The same holds for Yukawa coupling corrections \cite{yuk}. 
It turned out that these corrections can be quite large, up to 
$\pm 10\%$ for squark production, 
depending on the properties of the charginos, neutralinos, 
Higgs bosons, and squarks in the loops.  
In the remaining part of this section we will, however, not include Yukawa 
coupling corrections because they depend on the whole MSSM spectrum. 

\noi
Figure~\ref{fig:Xsectsl} shows the cross sections for $\stau$ and 
$\snu_\tau$ production for $\mstau{1}=156$ GeV, $\mstau{2}=180$ GeV, 
$\cstau=0.77$, and $m_{\snu_\tau}=148$ GeV. 
As can be seen, these cross sections can be comparable in size to 
$\st_1\bar{\st_1}$ production. In Fig.~\ref{fig:Xsectsl} we 
have included only ISR. Yukawa coupling corrections are below 
the percent level for this choice of parameters 
and e.g., $M=200$~GeV, $\mu=800$~GeV, $m_A=300$~GeV, $\tan\b=4$. 
They can, however, go up to $\sim 5\%$ in certain parameter regions, 
especially for large $\tan\b$, see \cite{yuk}.

\begin{figure}[p]
\graph{0cm}{0cm}{-7mm}{!}{!}{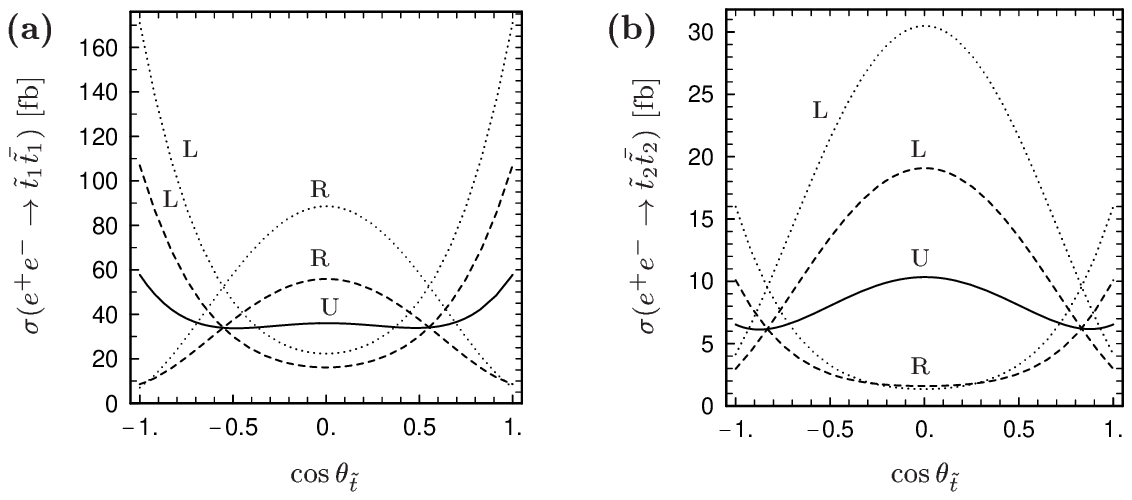}
\caption{$\cst$ dependence of stop pair production cross sections 
for $\mst{1}=200$ GeV, $\mst{2}=420$ GeV, $\msg=555$ GeV;  
$\sqrt{s}=500$ GeV in (a) and $\sqrt{s}=1$ TeV in (b) ; 
the label ``L'' (``R'') denotes $\Pm=-0.9$ (0.9) with 
the dashed lines for $\Pp=0$, and the dotted lines for $|\Pp|=0.6$ 
$[{\rm sign}(\Pp)=-{\rm sign}(\Pm)]$; 
the full lines labeled ``U'' are for unpolarized beams ($\Pm=\Pp=0$).
\label{fig:stopcth}}
\end{figure}

Let us now turn to the dependence on the mixing angles and beam polarizations. 
In Fig.~\ref{fig:stopcth} we show $\s(e^+e^-\to\st_i\bar{\st_i})$ 
as a function of $\cst$ for $\mst{1}=200$ GeV, $\mst{2}=420$ GeV,  
$\msg=555$ GeV, $\sqrt{s}=500$ GeV in (a) and $\sqrt{s}=1$ TeV in (b).
The full lines are for unpolarized beams, the dashed lines are for a 90\% 
polarized $e^-$ beam, and the dotted ones for 90\% polarized $e^-$ 
and 60\% polarized $e^+$ beams. 
As one can see, beam polarization strengthens the $\cst$ 
dependence and can thus be essential for determining the mixing angle. 
Moreover, it can be used to enhance the signal and/or reduce the background. 

\begin{figure}[p]
\graph{0cm}{0cm}{-7mm}{!}{!}{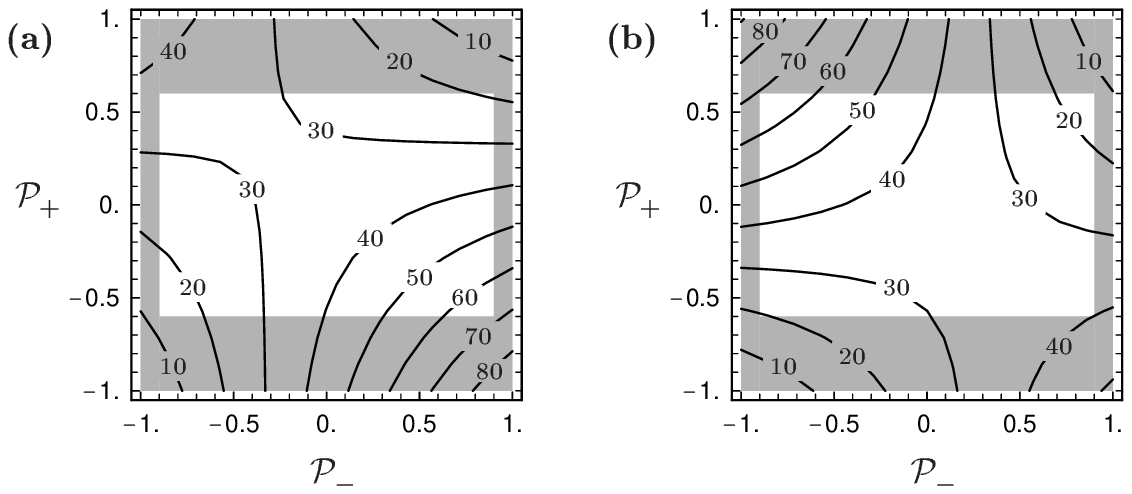}
\caption{Dependence of $\s(e^+e^-\to\st_1\bar{\st_1})$ on 
degree of electron and positron polarization for $\sqrt{s}=500$ GeV, 
$\mst{1}=200$ GeV, $\mst{2}=420$ GeV, and $\msg=555$ GeV; 
$\cst=0.4$ in (a) and $\cst=0.66$ in (b). 
\label{fig:stop11pol}}
\end{figure}

\noi
In Figs. \ref{fig:stop11pol}\,a,\,b we show the contour lines of
the cross section $\sigma (e^+e^- \ra \ti t_1 \bar{\ti t_1})$ as 
a function of the $e^-$ and $e^+$ beam polarizations $\cP_-$ and 
$\cP_+$ at $\sqrt{s}=500$~GeV for two values of $\cst$: $\cst = 0.4$ 
in (a) and $\cst= 0.66$ in (b). 
The white windows show the range of polarization of the TESLA 
design \cite{tesla}.   
As one can see, one can significantly increase the cross section by 
using the maximally possible $e^-$ {\it and} $e^+$ polarization. 
Here note that the (additional) positron polarization leads to an effective 
polarization \cite{baltay} of 
\beq
  \cP_{\rm eff}^{} = \frac{\cP_- - \cP_+}{1 - \cP_-\cP_+}\,.
\eeq

\begin{figure}[p]
\graph{0cm}{0cm}{-7mm}{!}{!}{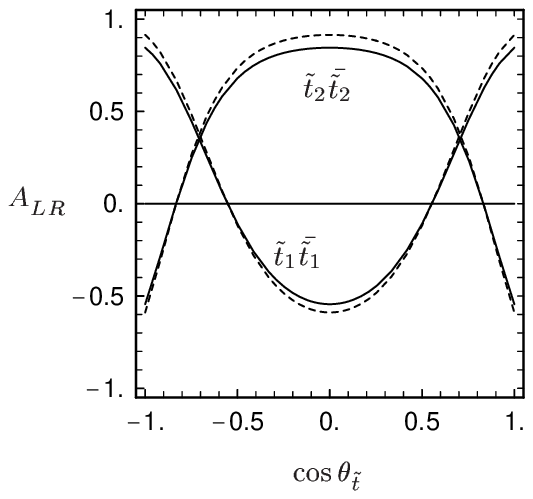}
\caption{$A_{LR}(e^+e^-\to\st_i\bar{\st_i})$ as function of $\cst$ for 
$\sqrt{s}=1$ TeV, $\mst{1}=200$ GeV, $\mst{2}=420$ GeV, and $\msg=555$ GeV; 
the solid lines are for 90\% polarized electrons and unpolarized positrons,
the dashed lines are for 90\% polarized electrons and 60\% polarized positrons.
\label{fig:stopALR}}
\end{figure}

\noi
In experiments with polarized beams one can also measure the 
left--right asymmetry  
\beq
  A_{LR} \equiv \frac{\s_L - \s_R}{\s_L+\s_R}
\label{eq:alr}
\eeq
where $\s_L^{} := \s\,(-|\cP_-|,|\cP_+|)$ and 
$\s_R^{} := \s\,(|\cP_-|,-|\cP_+|)$. 
This observable is sensitive to the amount of mixing of the 
produced sfermions while kinematical effects only enter at loop level.
In Fig.~\ref{fig:stopALR} we show $A_{LR}$ for  
$e^+ e^- \ra \st_i \bar{\st_i}$ ($i=1,2$)  
as a function of $\cst$ for 90\% polarized electrons
and unpolarized as well as 60\% polarized positrons;  
$\sqrt{s}=1$~TeV and the other parameters are as in Fig.~\ref{fig:stopcth}. 

\noi
Last but not least we note that the reaction $e^+e^-\to\st_1\st_2$ (with 
$\st_1\st_2\equiv \st_1\bar{\st_2}+ {\rm c.c.}$) 
can be useful to measure $\mst{2}$ below the $\st_2\bar{\st_2}$ threshold. 
As this reaction proceeds only via $Z$ exchange 
the cross section shows a clear $\sin 2\tst$ 
dependence. $\s(e^+e^-\to\st_1\st_2)$ can be enhanced by using 
left--polarized electrons. The additional use of right--polarized positrons 
further enhances the cross section. 
To give an example, for $\mst{1}=200$ GeV, $\mst{2}=420$ GeV, $\cst=0.7$, 
and $\sqrt{s}=800$ GeV we obtain $\s(\st_1\st_2)=7.9$\,fb, 9.1\,fb, 
and 14.1\,fb for 
$(\Pm,\,\Pp)=(0,\,0)$, $(-0.9,\,0)$, and $(-0.9,\,0.6)$, 
respectively. 
The left--right asymmetry, however, hardly varies with $\cst$: 
$A_{LR}(\st_1\st_2)\simeq 0.14$ (0.15) for $|\Pm|=0.9$ and $|\Pp|=0$ (0.6),  
$0<|\,\cst|<1$, and the other parameters as above.

\begin{figure}[p]
\graph{0cm}{-0.5cm}{-1.1cm}{!}{!}{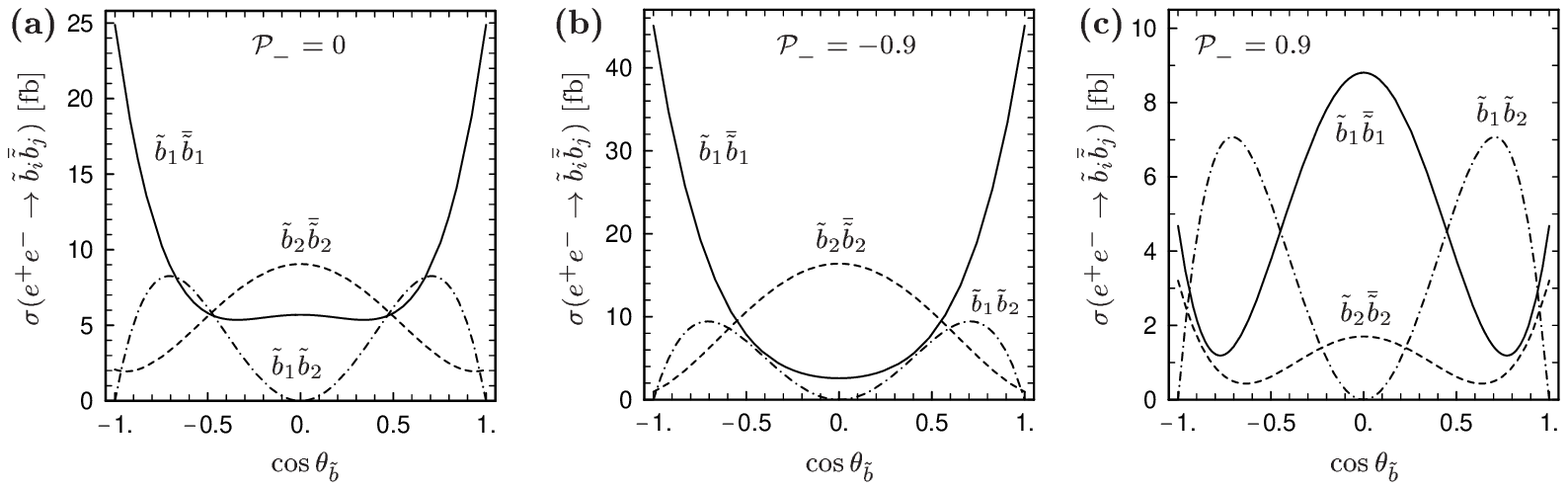}
\caption{$\csb$ dependence of sbottom pair production cross sections 
for $\msb{1}=284$ GeV, $\msb{2}=345$ GeV, $\msg=555$ GeV, 
$\sqrt{s}=800$ GeV and various $e^-$ beam polarizations ($\Pp=0$).
\label{fig:sbotcth}}
\end{figure}

We next discuss sbottom production using the scenario of 
Fig.~\ref{fig:Xsectsq}\,b, i.e. $\msb{1}=284$ GeV and $\msb{2}=345$ GeV. 
In this case, all three combinations $\sb_1\bar{\sb}_1$, 
$\sb_1\sb_2$ $(\equiv \sb_1\bar{\sb}_2+\sb_2\bar{\sb}_1)$, and 
$\sb_2\bar{\sb}_2$ can be produced at $\sqrt{s}=800$~GeV. 
The $\csb$ dependence of the corresponding cross sections 
are shown in Fig.~\ref{fig:sbotcth} for unpolarized, 90\% left--, 
and 90\% right--polarized electrons ($\cP_+ =0$). 
As can be seen, beam polarization can be a useful tool to disentangle 
$\sb_1$ and $\sb_2$.  

\begin{figure}[p]
\graph{0cm}{0cm}{-0.8cm}{!}{!}{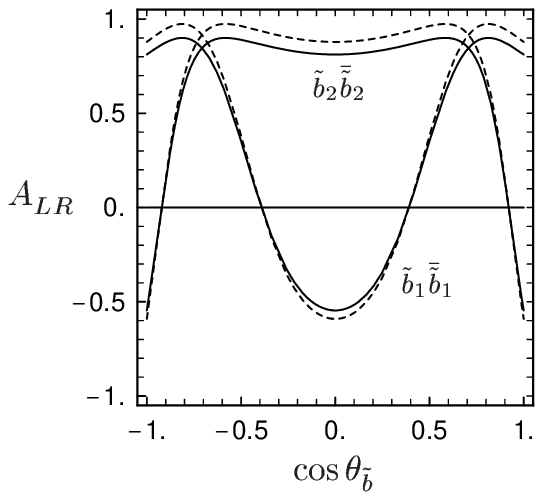}
\caption{$A_{LR}$ of sbottom pair production as function of $\csb$ 
for 90\% electron polarization; the full lines are for unpolarized 
and the dashed lines  for 60\% polarized positrons; $\sqrt{s}=800$ GeV, 
$\msb{1}=284$ GeV, $\msb{2}=345$ GeV, and $\msg=555$ GeV. 
\label{fig:sbotALR}}
\end{figure}

\noi
The left--right asymmetry $A_{LR}$, Eq.~(\ref{eq:alr}), 
of $\sb_1\bar{\sb}_1$ and $\sb_2\bar{\sb}_2$ production 
is shown in Fig.~\ref{fig:sbotALR} 
as a function of $\cos \tsb$ for 90\% polarized $e^-$ and 
unpolarized as well as 60\% polarized $e^+$ beams. 
As in the case of stop production, $A_{LR}(\sb_i\bar{\sb}_i)$ is 
very sensitive to the left--right mixing.

\begin{figure}[p]
\graph{0cm}{0cm}{-0.8cm}{!}{!}{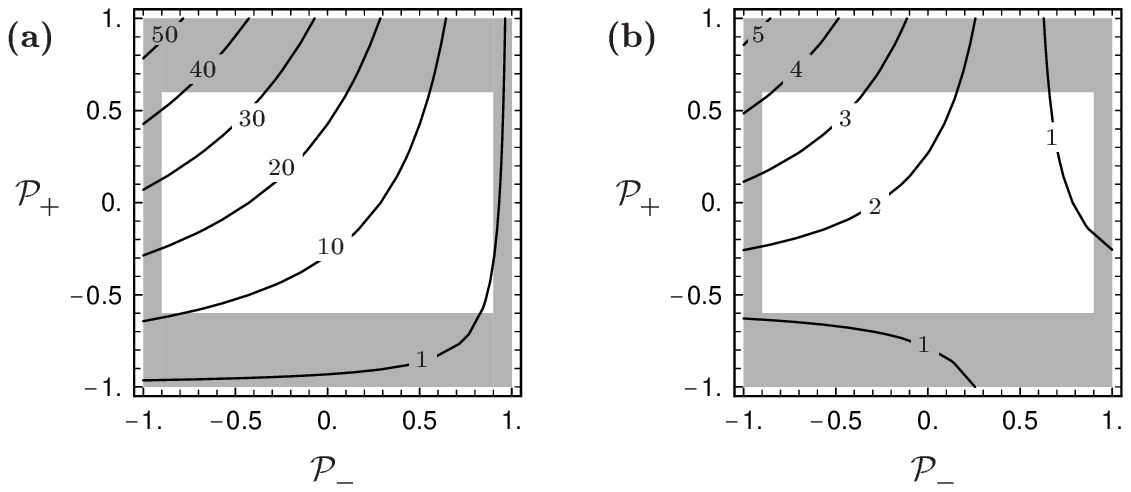}
\caption{Dependence of (a) $\s(e^+e^-\to\sb_1\bar{\sb_1})$ and 
(b) $\s(e^+e^-\to\sb_2\bar{\sb_2})$  (in fb) 
on the degree of electron and positron polarization for $\sqrt{s}=800$ GeV, 
$\msb{1}=284$ GeV, $\msb{2}=345$ GeV, $\csb=0.84$, and $\msg=555$ GeV.
\label{fig:sbotpol}}
\end{figure}

\noi
The explicit dependence on the $e^-$ and $e^+$ beam polarizations 
can be seen in Fig.~\ref{fig:sbotpol} where we plot the contourlines of 
$\sigma(e^+ e^- \ra \ti b_1 \bar{\ti b_1})$ and
$\sigma(e^+ e^- \ra \ti b_2 \bar{\ti b_2})$ as 
functions of $\cP_-$ and $\cP_+$ for the parameters used above  
and $\cos\tsb=0.84$. Again, the white windows indicate the
range of the TESLA design. 
Also in this case we observe that one can considerably increase 
the cross section by rising the effective polarization.

The renormalization group equations \cite{rge} 
for the slepton parameters are different from those for the squarks. 
Moreover, owing to Yukawa coupling effects, 
the parameters of the 3rd generation evolve differently compared to those 
of the 1st and 2nd generation. Therefore, measuring the properties 
of the squarks as well as the sleptons quite precisely will be necessary 
to test the 
boundary conditions at the GUT scale and the SUSY breaking mechanism. 

\begin{figure}[p]
\graph{0cm}{-0.5cm}{-1cm}{!}{!}{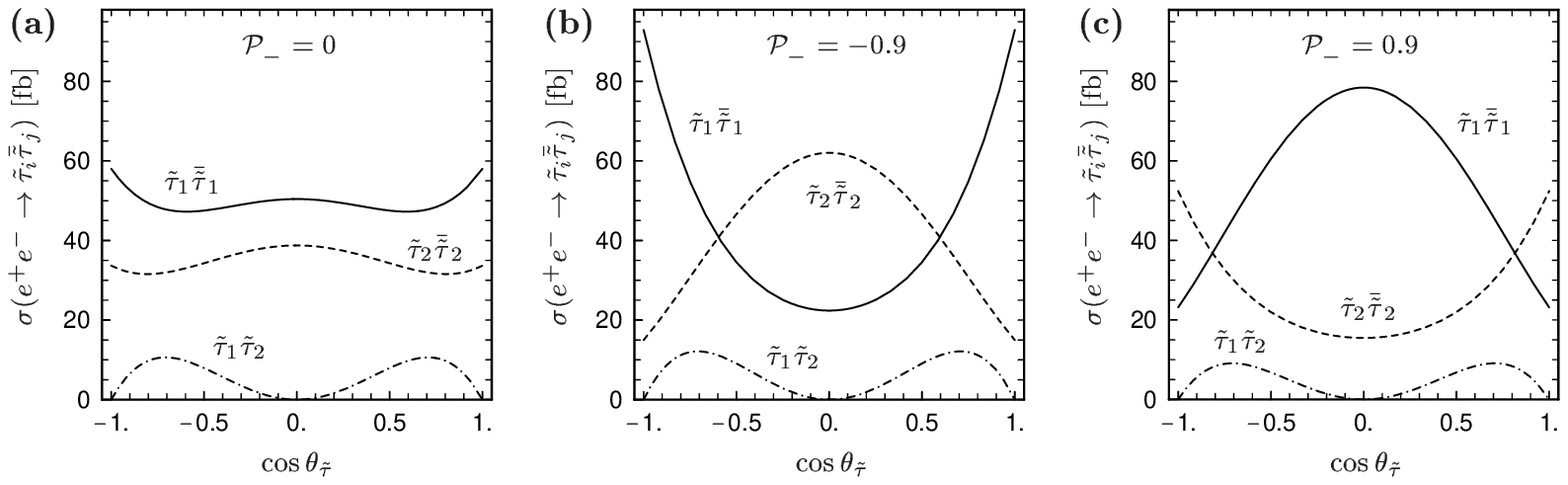}
\caption{$\cstau$ dependence of stau pair production cross sections 
for $\mstau{1}=156$ GeV, $\mstau{2}=180$ GeV, $\sqrt{s}=500$ GeV and 
various $e^-$ beam polarizations ($\Pp=0$).
\label{fig:staucth}}
\end{figure}

\begin{figure}[p]
\graph{0cm}{0cm}{-0.8cm}{!}{!}{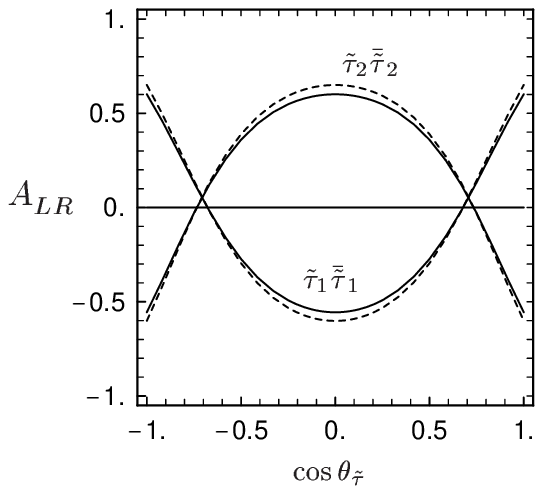}
\caption{$A_{LR}$ of stau pair production as function of $\cstau$ 
for 90\% electron polarization; the full lines are for unpolarized 
and the dashed lines  for 60\% polarized positrons; $\sqrt{s}=500$ GeV,    
$\mstau{1}=156$ GeV, $\mstau{2}=180$ GeV, and $\cstau=0.77$.
\label{fig:stauALR}}
\end{figure}

In the following plots on $\stau$ and $\snu_\tau$ pair production we fix 
$m_{\stau_1}=156$~GeV, $m_{\stau_2}=180$~GeV, and $m_{\snu}=148$~GeV  
as in Fig.~\ref{fig:Xsectsl}. 
In the calculation of the cross sections we include ISR 
corrections which turn out to be of the order of 10--15\%. 
Figure~\ref{fig:staucth} shows the $\cos \theta_{\stau}$ dependence 
of $\stau_i\bar{\stau}_j$ production at $\sqrt{s} = 500$~GeV for
unpolarized as well as for polarized $e^-$ beams ($\cP_+=0$). 
The usefulness of beam polarization to (i) increase the 
$\cos\t_{\stau}$ dependence and (ii) enhance/reduce $\stau_1\bar{\stau}_1$ 
relative to $\stau_2\bar{\stau}_2$ production is obvious. 
The left--right asymmetry of $\stau_i\bar{\stau}_i$ production 
for the parameters of Fig.~\ref{fig:staucth} is shown in 
Fig.~\ref{fig:stauALR}. Here note that, in contrast to $\st$ and $\sb$ 
production, $A_{LR}(\stau_i\bar{\stau}_i)$ is almost zero for 
maximally mixed staus.

\begin{figure}[p]
\graph{0cm}{-0.5cm}{-1.3cm}{!}{!}{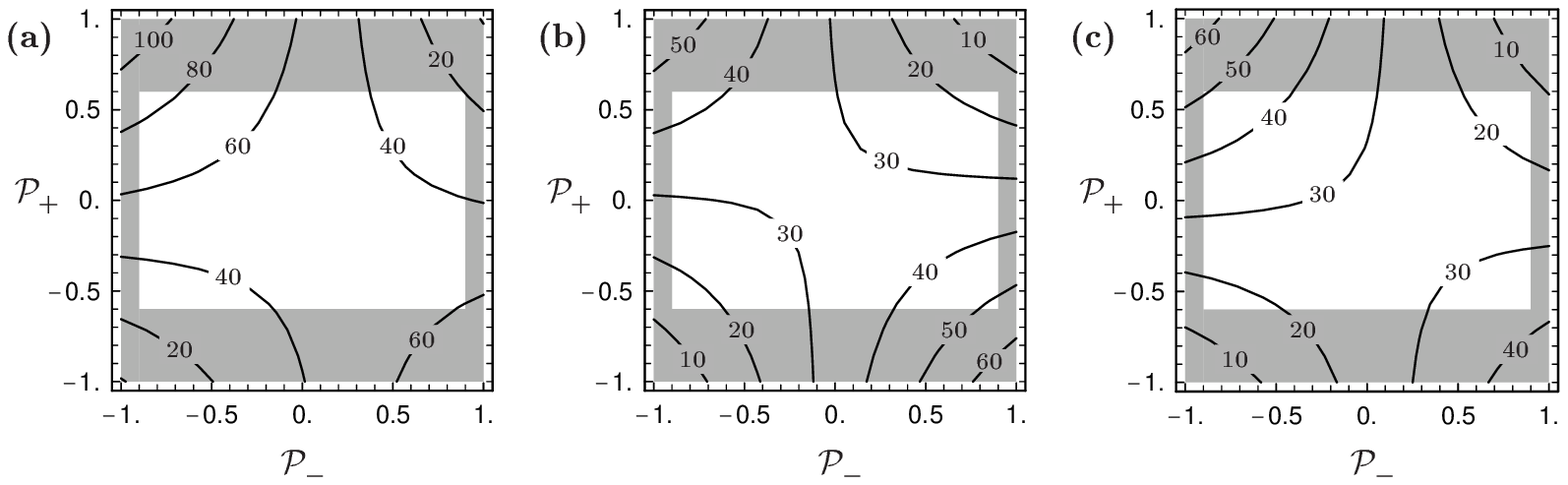}
\caption{Dependence of (a) $\s(e^+e^-\to\stau_1\bar\stau_1)$, 
(b) $\s(e^+e^-\to\stau_2\bar\stau_2)$,  and 
(c) $\s(e^+e^-\to\snu_\tau\bar\snu_\tau)$  (in fb)
on the degree of electron and positron polarization 
for $\sqrt{s}=500$ GeV, $\mstau{1}=156$ GeV, $\mstau{2}=180$ GeV, 
$\cstau=0.77$, and $m_{\snu_\tau}=148$ GeV.
\label{fig:staupol}}
\end{figure}

\noi
Finally, the dependence of $\s(e^+e^-\to\stau_i\bar{\stau}_i)$, 
for $\cth_{\stau}=0.77$, and $\s(e^+e^-\to\snu_\tau\bar{\snu}_\tau)$ 
on both the electron and positron polarizations is shown in 
Fig.~\ref{fig:staupol}. Notice that one could again substantially 
increase the cross sections by going beyond 60\% $e^+$ polarization. 

\section{Decays}

Owing to the influence of the Yukawa terms and
the left--right mixing, the decay patterns of stops,
sbottoms, $\tau$-sneutrinos, and staus are in general more
complicated than 
those of the sfermions of the first two generations.
As for the sfermions of the first and second generation, 
there are the decays into neutralinos or charginos ($i,j=1,2;~k=1,...4$): 
\beqa
  \st_i \;\to\; t\,\nt_k\,,\; b\,\ch^+_j, \;\;
  &\quad& 
  \sb_i \;\to\; b\,\nt_k\,,\; t\,\ch^-_j, \label{eq:decnc}
  \\
  \stau_i \;\,\to\; \tau\,\nt_k\,,\; \nu_\tau\,\ch^-_j, 
  &\quad& 
  \snu_\tau \,\to\; \nu_\tau\,\nt_k\,,\; \tau\,\ch^+_j. 
\eeqa
Stops and sbottoms may also decay into gluinos, 
\beq
  \st_i \ra t\,\sg\,, \qquad \sb_i \ra b\,\sg 
\eeq
and if these decays are kinematically allowed, they are important. 
If the mass differences $|\mst{i} - \msb{j}|$ and/or  
$|m_{\stau_i} - m_{\snu_\tau}|$ are large enough the transitions
\beq
  \st_i \ra \sb_j\, + W^+\,(H^+) \quad {\rm or} \quad
  \sb_i \ra \st_j\, + W^-\,(H^-) \label{eq:sqWH}
\eeq
as well as
\beq
  \stau_i   \ra \snu_\tau + W^- \, (H^-)   \quad {\rm or} \quad
  \snu_\tau \ra \stau_i   + W^+ \, (H^+) \label{eq:slWH}
\eeq
can occur.  
Moreover, in case of strong $\sf_L^{}-\sf_R^{}$ mixing the
splitting between the two mass eigenstates may be so large that
heavier sfermion can decay into the lighter one:
\beq
  \st_2 \to \st_1 + Z^0\,(h^0, H^0, A^0)\,, \quad
  \sb_2 \to \sb_1 + Z^0\,(h^0, H^0, A^0)\,, \quad
  \stau_2 \to \stau_1 + Z^0\,(h^0, H^0, A^0)\,. \label{eq:decZH}
\eeq
The SUSY--QCD corrections to the squark decays of 
Eqs.~(\ref{eq:decnc}) to (\ref{eq:decZH}) have 
been calculated in \cite{deccor}. 
The Yukawa coupling corrections to the decay $\sb_i\to t\ch^-_j$ 
have been discussed in \cite{ghs}. 
All these corrections will be important for precision measurements. 

The decays of the lighter stop can be still more complicated:
If $\mst{1} < \mnt{1}+m_t$ and $\mst{1} < \mch{1}+m_b$ the 
three--body decays \cite{werner3bdy}
\beq  
  \st_1\;\to\; W^+\,b\,\nt_1 , \quad H^+\,b\,\nt_1 , 
         \quad b\,{\tilde l}^+_i \, \nu_l , 
         \quad b\,{\tilde \nu}_l \, l^+ \label{eq:3bdy}
\eeq
can compete with the loop--decay \cite{hikasa-kobayashi}
\beq 
  {\tilde t}_1 \to c \, {\tilde \chi}^0_{1,2} . 
\eeq
If also $m_{\ti t_1} < m_{\ti \chi^0_1} + m_b + m_W$ etc., then 
four--body decays \cite{djouadi4bdy} 
\beq 
  \ti t_1 \ra b\, f\bar{f'}\, \nt_1
\eeq 
have to be taken into account.

\begin{figure}[t]
\graph{0cm}{0cm}{-0.6cm}{!}{!}{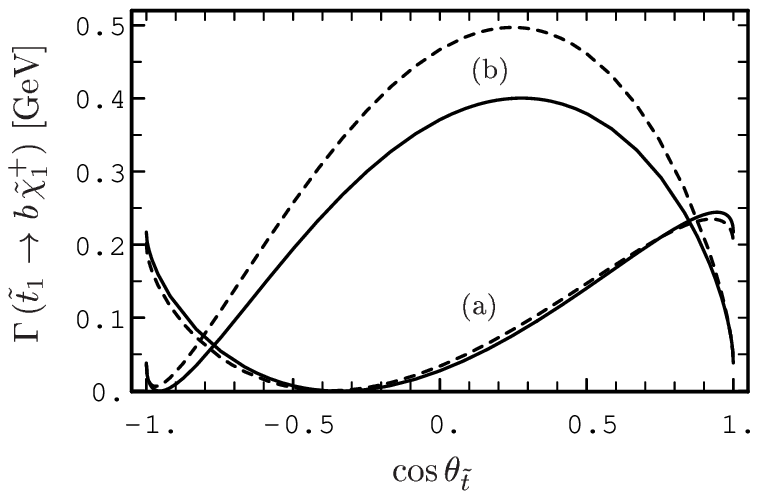}
\caption{Decay width of $\st_1$ as a function of $\cst$ for 
$\mst{1}=200$ GeV, $\mst{2}=420$ GeV, $\tan\b=4$, 
$\{M,\,\mu\}=\{180,\,360\}$ GeV in (a), and 
$\{M,\,\mu\}=\{360,\,180\}$ GeV in (b); the dashed lines show the results 
at tree level, the full lines those at ${\cal O}(\a_s)$.
\label{fig:gamst1}}
\end{figure}

We have studied numerically the widths and branching ratios
of the various sfermion decay modes. In the calculation of the
stop and sbottom decay widths we have included the SUSY--QCD 
corrections according to \cite{deccor}. 
In Fig.~\ref{fig:gamst1} we show the decay width of
$\st_1 \to b \ti \chi^+_1$ as a function of $\cos\tst$ for 
$m_{\ti t_1} = 200$~GeV, $m_{\ti t_2} = 420$~GeV, $\tan \beta = 4$,  
$M=180$~GeV, $\mu = 360$~GeV, (gaugino--like $\ti \chi^+_1$), 
and $M=360$~GeV, $\mu = 180$~GeV (higgsino--like $\ti \chi^+_1$). 
If $\Gamma (\ti t_1)\lsim 200$~MeV $\st_1$ may hadronize 
before decaying \cite{drees-eboli}. 
In Fig.~\ref{fig:gamst1} this is the case for a gaugino--like $\ch^+_1$ 
as well as for a higgsino--like one if $\cos\tst \lsim -0.5$ 
(or $\cos\tst \gsim 0.9$).

In case that all tree--level two--body decay modes are forbidden for $\st_1$,  
higher order decays are important for its phenomenology. In the following
we study examples where three--body decay modes, 
Eq.~(\ref{eq:3bdy}), are the dominant ones.
For fixing the parameters we choose the following
procedure: in addition to $\tan \beta$ and $\mu$ we use 
$m_{\tilde{t}_{1}}$ and $\cos \theta_{\tilde{t}}$ as input parameters 
in the stop sector.
For the sbottom (stau) sector we fix
$M_{\ti Q}, M_{\ti D}$ and $A_b$ ($M_E, M_L, A_\tau$) as input parameters.
(We use this mixed set of parameters in order to avoid unnaturally 
large values for $A_b$ and $A_\tau$.) 
Moreover, we assume for simplicity that the soft SUSY breaking parameters 
are equal for all generations. 
Note that, because of SU(2) invariance $M_{\ti Q}$ appears in both 
the stop and sbottom mass matrices, see Eqs.~(1)--(3). 
The mass of the heavier stop can thus be calculated
from the above set of input parameters as: 
\begin{equation}
  m^2_{\tilde{t}_{2}} = 
  \frac{2 M^2_{\ti Q} + 2\, m^2_Z \cos 2 \beta 
    \left(\frac{1}{2} - \frac{2}{3} \sin^2 \theta_{W} \right)
    + 2\,m^2_t - m^2_{\tilde{t}_{1}} (1 + \cos 2 \theta_{\tilde{t}} ) }
  { 1 - \cos 2 \theta_{\tilde{t}} }
\end{equation}
In the sbottom (stau) sector obviously the physical quantities $m_{\tilde{b}_{1}}$,
$m_{\tilde{b}_{2}}$, and $\cos \theta_{\tilde{b}}$ 
($m_{\tilde{\tau}_{1}}, m_{\tilde{\tau}_{2}}$, $\cos \theta_{\tilde{\tau}}$) 
change with $\mu$ and $\tan \beta$.

\begin{figure}[h!]
\graph{0cm}{0cm}{-0.6cm}{!}{!}{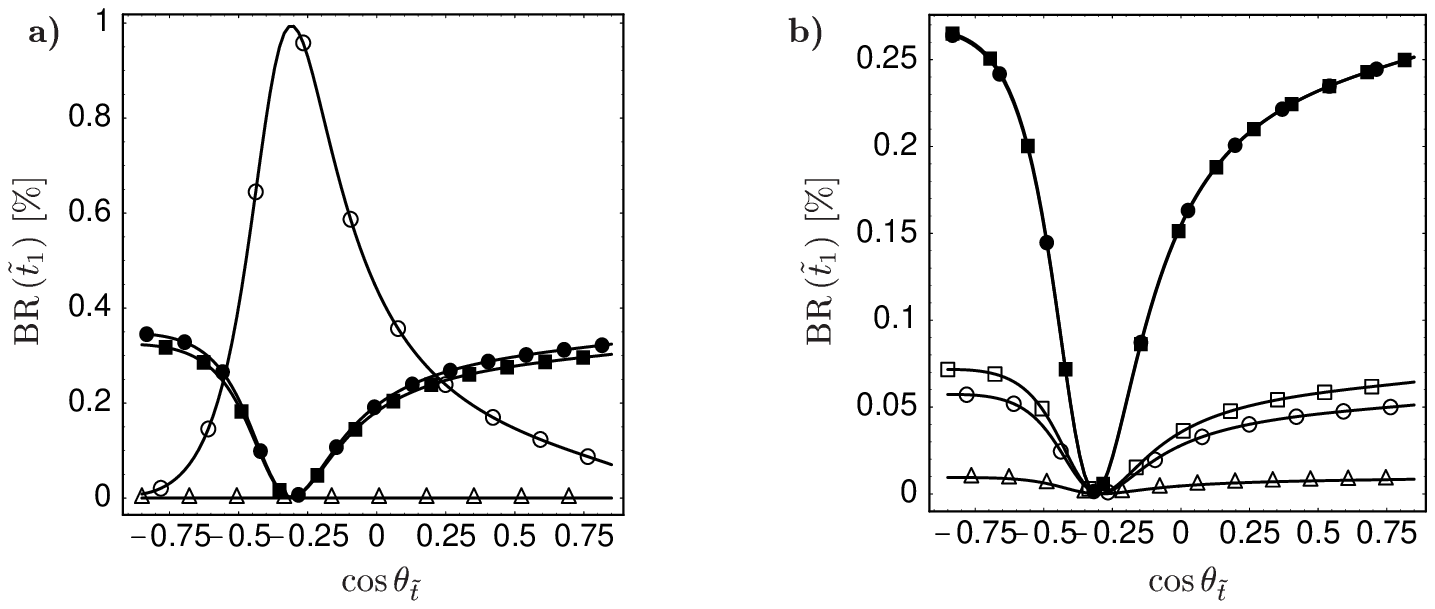}
\caption[]{Branching ratios for $\tilde{t}_{1}$ decays as a function of $\cos \theta_{\tilde{t}}$
            for $m_{\tilde{t}_{1}} = 220$~GeV, $\tan \beta = 3$, $\mu = 500$~GeV, and
            $M = 240$~GeV. The other parameters are given in 
            Table~\ref{tab:brst3}.
           The curves in a) correspond to the transitions:
           $\circ \hspace{1mm} \tilde{t}_{1} \to b \, W^+ \, \tilde{\chi}^0_{1}$,
           $\triangle \hspace{1mm} \tilde{t}_{1} \to c \tilde{\chi}^0_{1}$, 
           ~\recht $(\tilde{t}_{1} \to b \, e^+ \, \tilde{\nu}_e)$
                + $(\tilde{t}_{1} \to b \, \nu_e \, \tilde{e}^+_L)$, and
           $\bullet \hspace{1mm} (\tilde{t}_{1} \to b \, \tau^+ \, \tilde{\nu}_{\tau})$
                + $(\tilde{t}_{1} \to b \, \nu_\tau \, \tilde{\tau}_{1})$
                + $(\tilde{t}_{1} \to b \, \nu_\tau \, \tilde{\tau}_{2})$.
           The curves in b) correspond to the transitions:
           $\circ \hspace{1mm} \tilde{t}_{1} \to b \, \nu_e \, \tilde{e}^+_L$, 
           \rechtl $\tilde{t}_{1} \to b \, \nu_\tau \, \tilde{\tau}_{1}$,
           $\triangle \hspace{1mm} \tilde{t}_{1} \to b \, \nu_\tau \, \tilde{\tau}_{2}$,
           \recht $\tilde{t}_{1} \to b \, e^+ \, \tilde{\nu}_e$, and
           $\bullet \hspace{1mm} \tilde{t}_{1} \to b \, \tau^+ \, \tilde{\nu}_{\tau}$.
\label{fig:brst3cosa}      }
\end{figure}

\noi
A typical example is given in Fig.~\ref{fig:brst3cosa} where we show the branching
ratios of $\st_1$ as a function of $\cos \theta_{\tilde{t}}$. 
We have restricted the $\cos \theta_{\tilde{t}}$ range in such
a way that $|A_t| \lsim 1$~TeV to avoid color/charge breaking minima. 
The parameters and physical quantities are given in Table~\ref{tab:brst3}. 
In Fig.~\ref{fig:brst3cosa}\,a we present 
BR$(\tilde{t}_{1} \to b \, W^+ \, \tilde{\chi}^0_{1})$,
BR$(\tilde{t}_{1} \to c \, \tilde{\chi}^0_{1})$,
BR$(\tilde{t}_{1} \to b \, e^+ \, \tilde{\nu}_e$) + 
BR$(\tilde{t}_{1} \to b \, \nu_e \, \tilde{e}^+_L)$, and
BR$(\tilde{t}_{1} \to b \, \tau^+ \, \tilde{\nu}_{\tau})$ +
BR$(\tilde{t}_{1} \to b \, \nu_\tau \, \tilde{\tau}_{1})$ +
BR$(\tilde{t}_{1} \to b \, \nu_\tau \, \tilde{\tau}_{2})$. 
Here the decay into $b\,H^+\,\tilde{\chi}^0_{1}$ is not included because 
for the parameters of Table~\ref{tab:brst3} there is no $m_A$ which 
simultaneously allows this decay and fulfils the condition 
$m_{h^0} \gsim 90$~GeV. However, in general this decay is suppressed
by kinematics \cite{werner3bdy}.
We have summed the branching ratios of those decays which give the 
same final states 
after the sleptons have decayed. For example:
\begin{eqnarray}
\tilde{t}_{1} \to b \, \nu_\tau \, \tilde{\tau}_{1} \,
          \to \, b \, \tau \, \nu_\tau \, \tilde{\chi}^0_{1}, \hspace{5mm}
\tilde{t}_{1} \to b \, \tau \, \tilde{\nu}_{\tau} \,
          \to \, b \, \tau \, \nu_\tau \, \tilde{\chi}^0_{1}.
\end{eqnarray}
Note, that the requirement $m_{\tilde{t}_{1}} - m_b < m_{\tilde{\chi}^+_1}$
implies that the sleptons can only decay into the corresponding lepton plus
the lightest neutralino except for a small parameter region where
the decay into $\tilde{\chi}^0_{2}$ is possible. However, there this decay is 
negligible due to kinematics. The branching ratios for decays into
$\tilde{\mu}_L^{}$ or $\tilde{\nu}_{\mu}$ are not shown because they are the
same as those of the decays into $\tilde{e}_L^{}$ or $\tilde{\nu}_{e}$ 
up to very tiny mass effects. 
The sum of the branching ratios for the decays into $\tilde{\tau}_{1}$ and 
$\tilde{\tau}_{2}$ also has nearly the same size as $\tan \beta$ is small. 
BR$(\tilde{t}_{1} \to c \, \tilde{\chi}^0_{1})$ is of order $10^{-4}$ 
independent of $\cos \theta_{\tilde{t}}$ and therefore 
negligible. Near $\cos \theta_{\tilde{t}} = -0.3$ the decay into 
$b \, W^+ \, \tilde{\chi}^0_{1}$ has a branching ratio of $\sim 100\%$. 
Here the $\st_1 b \ch^+_1$ coupling vanishes leading to the
reduction of the decays into sleptons.

\begin{table}[t]
\begin{center}
\begin{tabular}{|c|c|c||c|c|c|}
\hline 
$\tan \beta$ & $\mu$ & $M$ & $m_{\tilde{\chi}^0_1}$ 
& $m_{\tilde{\chi}^+_1}$ & $m_{\tilde{\chi}^+_2}$ \\ \hline
3 & 500 & 240 &  116 & 223 & 520 \\ \hline \hline
$M_D$ & $M_Q$ & $A_b$  & $m_{\tilde{b}_{1}}$ & $m_{\tilde{b}_{2}}$ 
& $\cos \theta_{\tilde{b}}$ \\ \hline
370 & 340 & 150 & 342 & 372 & 0.98 \\ \hline \hline
$M_E$ & $M_L$ & $A_\tau$  & $m_{\tilde{\tau}_{1}}$ & $m_{\tilde{\tau}_{2}}$ 
& $\cos \theta_{\tilde{\tau}}$ \\ \hline
190 & 190 & 150 & 188 & 200 & 0.69 \\ \hline \hline
 & & $m_{\tilde{t}_{1}}$ &
            $m_{\tilde{e}_{L}}$ & $m_{\tilde{e}_{R}}$ & $m_{\tilde{\nu}_{\tau}}$
\\ \hline
 & & 220 & 195 & 195 & 181 \\ \hline
\end{tabular}
\end{center}
\caption[]{Parameters and physical quantities used in Fig.~\ref{fig:brst3cosa}
           and \ref{fig:brst3tan}.
            All masses are given in GeV.
        \label{tab:brst3}   }
\end{table}

\noi
In Fig.~\ref{fig:brst3cosa}b the branching ratios for the decays into the
different sleptons are shown. 
As $\tan \beta$ is small the sleptons couple mainly to the gaugino components 
of $\tilde{\chi}^+_{1}$. 
Therefore, the branching ratios of decays into staus, which are strongly 
mixed, are reduced.
However, the sum of both branching ratios is
nearly the same as BR$(\tilde{t}_{1} \to b \, \nu_e \, \tilde{e}^+_L)$.
The decays into sneutrinos are preferred by kinematics. 
The decay $\st_1\to b \, W^+ \, \tilde{\chi}^0_{1}$ is dominated 
by top quark exchange, followed by chargino contributions. 
In many cases the interference term between $t$ and $\tilde{\chi}^+_{1,2}$ 
is more important than the pure $\tilde{\chi}^+_{1,2}$ exchange. 
Moreover, we have found that the contribution from sbottom exchange is 
in general negligible.

\begin{figure}[h!]
\graph{0cm}{0cm}{-0.7cm}{!}{!}{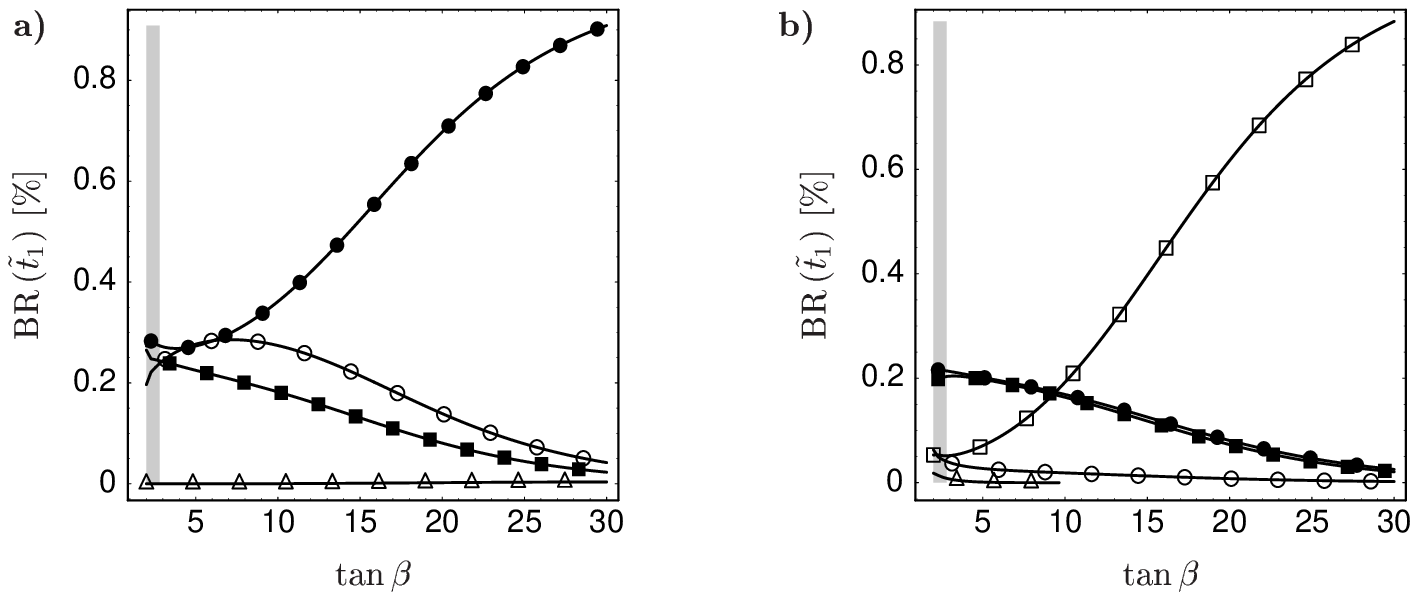}
\caption[]{Branching ratios for $\tilde{t}_{1}$ decays as a function of $\tan \beta$
            for $m_{\tilde{t}_{1}} = 220$~GeV, $\mu = 500$~GeV, $\cos \theta_{\tilde{t}} = 0.25$ and
            $M = 240$~GeV. The other parameters are given in 
            Table~\ref{tab:brst3}.
           The curves in a) correspond to the transitions:
           $\circ \hspace{1mm} \tilde{t}_{1} \to b \, W^+ \, \tilde{\chi}^0_{1}$,
           $\triangle \hspace{1mm} \tilde{t}_{1} \to c \tilde{\chi}^0_{1}$, 
           ~\recht $(\tilde{t}_{1} \to b \, e^+ \, \tilde{\nu}_e)$
                + $(\tilde{t}_{1} \to b \, \nu_e \, \tilde{e}^+_L)$, and
           $\bullet \hspace{1mm} (\tilde{t}_{1} \to b \, \tau^+ \, \tilde{\nu}_{\tau})$
                + $(\tilde{t}_{1} \to b \, \nu_\tau \, \tilde{\tau}_{1})$
                + $(\tilde{t}_{1} \to b \, \nu_\tau \, \tilde{\tau}_{2})$.
           The curves in b) correspond to the transitions:
           $\circ \hspace{1mm} \tilde{t}_{1} \to b \, \nu_e \, \tilde{e}^+_L$, 
           \rechtl $\tilde{t}_{1} \to b \, \nu_\tau \, \tilde{\tau}_{1}$,
           $\triangle \hspace{1mm} \tilde{t}_{1} \to b \, \nu_\tau \, \tilde{\tau}_{2}$,
           \recht $\tilde{t}_{1} \to b \, e^+ \, \tilde{\nu}_e$, and
           $\bullet \hspace{1mm} \tilde{t}_{1} \to b \, \tau^+ \, \tilde{\nu}_{\tau}$. In
           gray area $m_{h^0} < 90$~GeV.
\label{fig:brst3tan}      }
\end{figure}

\noi
In Fig.~\ref{fig:brst3tan} we show the branching ratios of $\st_1$ decays 
as a function of $\tan \beta$
for $\cos \theta_{\tilde{t}} = 0.6$ and the other parameters as above. 
For small $\tan \beta$ the decay 
$\tilde{t}_{1} \to b \, W^+ \, \tilde{\chi}^0_{1}$ is the most important one.
The branching ratios for the decays into sleptons are reduced in the range 
$\tan \beta \lsim 5$ because the gaugino component of $\tilde{\chi}^+_{1}$ 
decreases and its mass increases.
For $\tan \beta \gsim 10$ the decays into the
$b \, \tau \, E \hspace{-2.5mm} / \hspace{1.2mm}$ final state become
more important because of the growing $\tau$ Yukawa coupling and because of
kinematics ($m_{\tilde{\tau}_{1}}$ decreases with increasing $\tan \beta$). 
Here $\tilde{t}_{1} \to b \, \nu_\tau \, \tilde{\tau}_{1}$
gives the most important contribution as can be seen in Fig.~\ref{fig:brst3tan}b.
Even for large $\tan \beta$ the decay into $c \, \tilde{\chi}^0_{1}$ 
is always suppressed.

\noi
From the requirement that no two--body decays be allowed at
tree level follows that $m_{\tilde{\chi}^+_1} > m_{\tilde{t}_{1}} - m_b$. 
Therefore, one expects an increase of 
BR$(\tilde{t}_{1} \to b \, W^+ \, \tilde{\chi}^0_{1})$ if $m_{\tilde{t}_{1}}$ 
increases,  because the decay into $b \, W^+ \, \tilde{\chi}^0_{1}$ 
is dominated by top--quark exchange
whereas for the decays into sleptons the $\tilde{\chi}^+_1$ contribution is
the dominating one. 
In general BR$(\tilde{t}_{1} \to b \, W^+ \, \tilde{\chi}^0_{1})$ is larger 
than $80\%$ if $m_{\tilde{t}_{1}} \gsim 350$~GeV \cite{werner3bdy}.


If the three--body decay modes are kinematically forbidden (or suppressed) 
four--body decays $\st_1\to b\,f\bar f\,\nt_1$ come into play. 
Depending on the MSSM parameter region, these decays can also dominate 
over the decay into $c\,\nt_1$. 
For a discussion, see \cite{djouadi4bdy}.

\begin{figure}[t]
\graph{0cm}{-0.3cm}{-0.7cm}{!}{!}{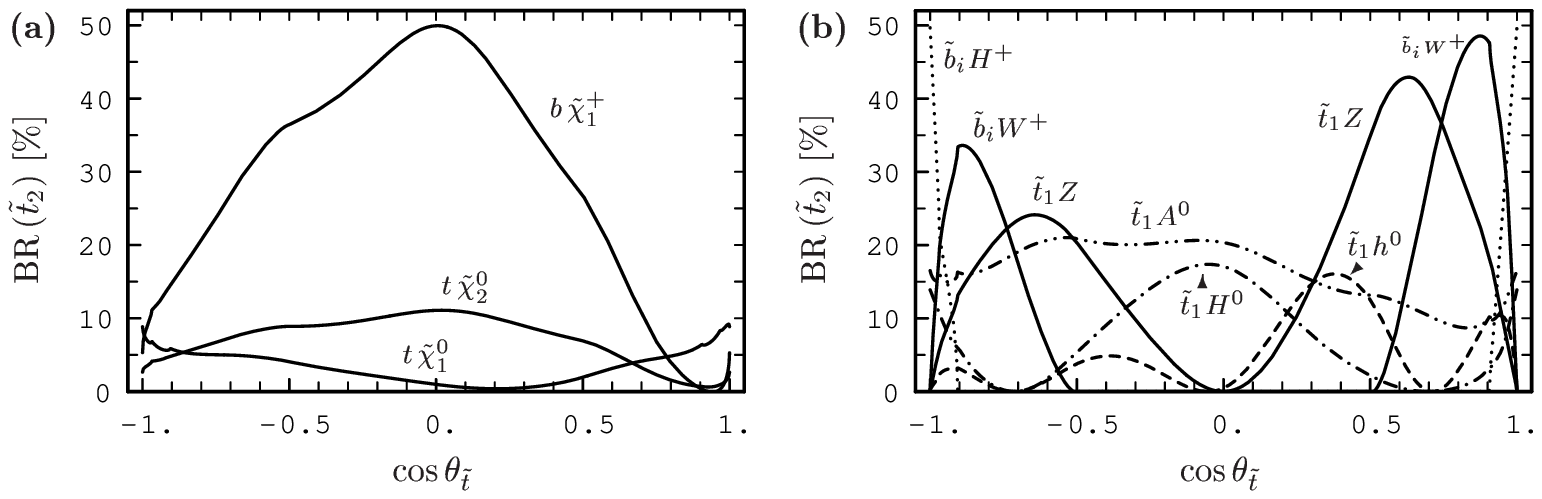}
\caption{Branching ratios of $\st_2$ decays at ${\cal O}(\a_s)$ 
as a function of $\cst$ for 
$\mst{1}=200$ GeV, $\mst{2}=420$ GeV, $M=180$ GeV, $\mu=360$ GeV, 
$\tan\b=4$, $M_{\ti D}=1.1\,M_{\ti Q}$, $A_b=-300$ GeV, and $m_A=200$ GeV; 
the fermionic decays are shown in (a) and the bosonic ones in (b).
\label{fig:BRst2-2}}
\end{figure}
\begin{figure}[t]
\graph{0cm}{-0.3cm}{-1.1cm}{!}{!}{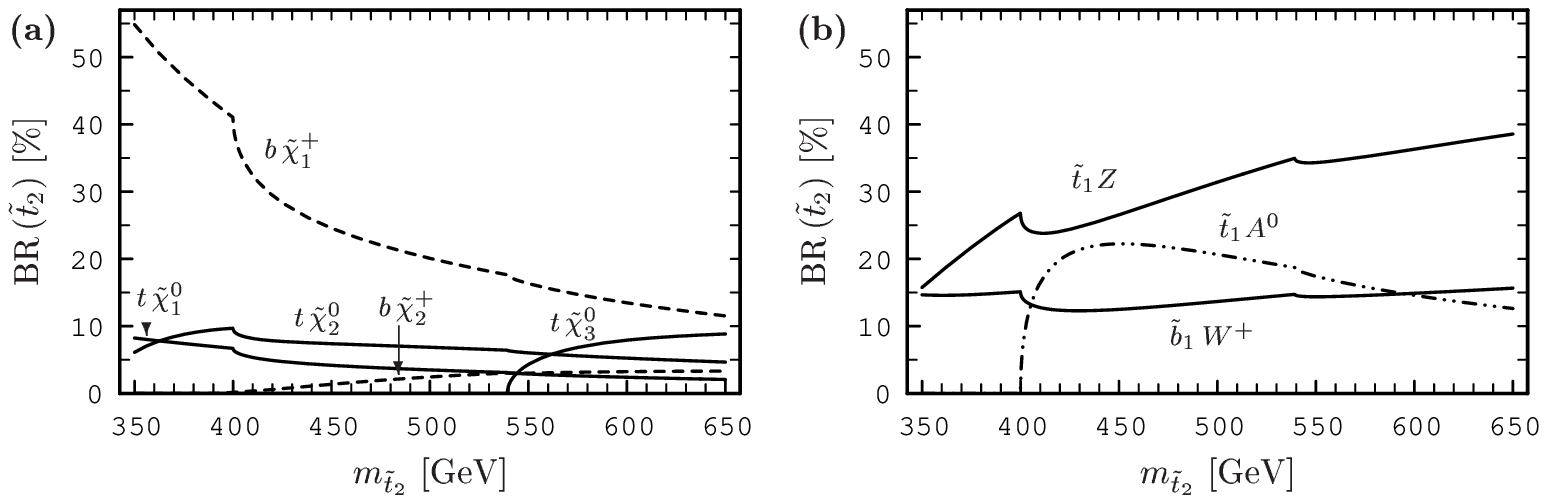}
\caption{Branching ratios of $\st_2$ decays at ${\cal O}(\a_s)$ 
as a function of $\mst{2}$ for 
$\mst{1}=200$ GeV, $\cst=-0.66$, $M=180$ GeV, $\mu=360$ GeV, 
$\tan\b=4$, $M_{\ti D}=1.1\,M_{\ti Q}$, $A_b=-300$ GeV, and $m_A=200$ GeV; 
the fermionic decays are shown in (a) and the bosonic ones in (b).
\label{fig:BRst2-3}}
\end{figure}

We now turn to the decays of $\st_2$. 
Here the bosonic decays of Eqs.~\ref{eq:sqWH} and \ref{eq:decZH} can 
play an important r\^{o}le as demonstrated in Figs.~\ref{fig:BRst2-2} 
and \ref{fig:BRst2-3}. In Fig.~\ref{fig:BRst2-2} we show 
the $\cst$ dependence of BR$(\st_2)$ for 
$\mst{1}=200$ GeV, $\mst{2}=420$ GeV, $M=180$ GeV, $\mu=360$ GeV, 
$\tan\b=4$, $M_{\ti D}=1.1\,M_{\ti Q}$, $A_b=-300$ GeV, and $m_A=200$ GeV.  
As can be seen, the decays into bosons can have branching ratios 
of several ten percent. 
The branching ratio of the decay into the gaugino--like $\ch^+_1$ 
is large if $\st_2$ has a rather strong $\st_L$ component. 
The decay $\st_2\to\st_1 Z$ is preferred by strong mixing. 
The decays into $\sb_i W^+$ only occur for $|\cos\tst|\gsim 0.5$ 
because the sbottom masses are related to the stop parameters 
by our choice $M_{\ti D}=1.1\,M_{\ti Q}$. 
Notice that BR($\st_2\to \sb_i W^+$) goes to zero for $|\cos\tst|= 1$ 
as in this case $\st_2=\st_R$. 
We have chosen $m_A$ such that decays into all MSSM Higgs bosons be  
possible. These decays introduce a more complicated dependence 
on the mixing angle, Eq.~(\ref{eq:mixangle}), because 
$A_t=(\mst{1}^2-\mst{2}^2)/(2m_t)+\mu\cot\b$ 
directly enters the stop--Higgs couplings.  
Here notice also the dependence on the sign of $\cos\tst$. 

\noi
Figure~\ref{fig:BRst2-3} shows the branching ratios of $\st_2$ decays 
as a function of $\mst{2}$ for $\cst=-0.66$ and the other parameters as 
in Fig.~\ref{fig:BRst2-2}. Again we compare the fermionic and bosonic 
decay modes. While for a rather light $\st_2$ the decay into 
$b\ch^+_1$ is the most important one, with increasing mass difference 
$\mst{2}-\mst{1}$ the decays into bosons, especially 
$\st_2\to\st_1 Z$, become dominant. 

Concerning the decays of $\sb_1$ and $\sb_2$ 
we found that the allowed range of e.g., $\msb{i}$ or $\csb$ 
is very restricted, once the other parameters are fixed.   
Therefore, we do not show figures of sbottom branching ratios. 
In general, however, for $\sb_1$ the decays into $b\nt_1$ 
and $b\nt_2$ are important for gaugino--like $\nt_{1,2}$ 
because the decay $\sb_1\to t\ch^-_1$ is kinematically suppressed. 
For $\sb_2$, decays into $Z$ and/or neutral Higgs bosons are 
important if the mass difference $\msb{2}-\msb{1}$ is large enough. 
This may well be the case for large $\tan\b$ and/or large $\mu$.   
If strong mixing in the stop sector leads to a light $\st_1$ 
then also the decays $\sb_i\to \st_1W^-\,(H^-)$ can have large branching 
ratios. 
For $\msb{1}=284$ GeV, $\msb{2}=345$ GeV, $\csb=0.84$,  
$M=180$ GeV, $\mu=360$ GeV, and $\tan\b=10$ for instance  
(taking $M_{\ti U}=0.9\,M_{\ti Q}$ and $A_t=-375$ GeV to fix the stop sector),  
we find BR$(\sb_1\to b\nt_1)=16\%$, BR$(\sb_1\to b\nt_2)=58\%$, 
BR$(\sb_1\to \st_1W^-)=26\%$, and  
BR$(\sb_2\to b\nt_1)=10\%$, BR$(\sb_2\to b\nt_2)=15\%$, 
BR$(\sb_2\to \st_1W^-)=74\%$. 
More details and plots on stop and sbottom decays can be found in 
\cite{bamapo,sabine,sqbosdec,wernerdiss}. 

\begin{figure}[t]
\graph{0cm}{0cm}{-1.cm}{!}{!}{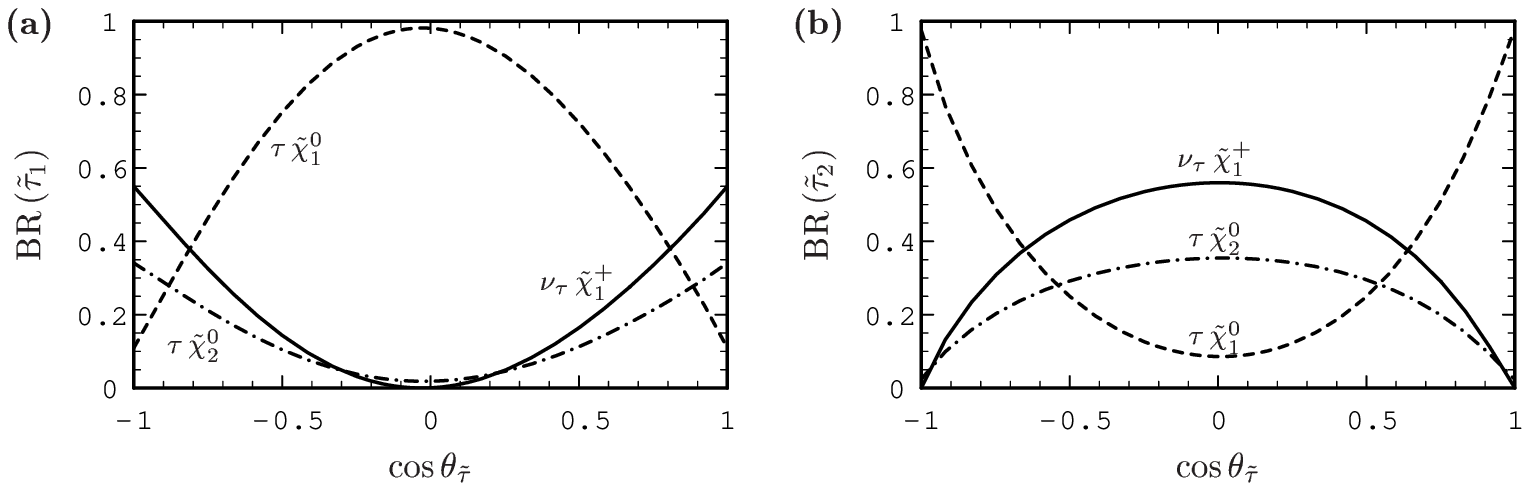}
\caption{Branching ratios of $\tilde\tau_1$ (a) and
$\tilde\tau_2$ (b) decays 
as a function of $\cos\theta_{\tilde\tau}$ for 
$m_{\tilde\tau_1}=156$ GeV, $m_{\tilde\tau_2}=180$ GeV, 
$M=120$~GeV, $\mu=300$~GeV, and $\tan\beta=4$.
\label{fig:BRstau1-2ct} }
\end{figure}

For the discussion of $\stau_{1,2}$ and $\snu_\tau$ decays we first consider  
the scenario of Fig.~\ref{fig:Xsectsl} where all particles are relatively 
light and have only fermionic decay modes. 
In Fig.~\ref{fig:BRstau1-2ct} we show the branching ratios of $\stau_1$ 
and $\stau_2$ decays as a function of $\cos\t_{\stau}$ for 
$m_{\tilde\tau_1}=156$ GeV, $m_{\tilde\tau_2}=180$ GeV, 
$M=120$~GeV, $\mu=300$~GeV, and $\tan\beta=4$.
In this case, decays into $\nt_1\sim\ti B$, $\nt_2\sim\ti W^3$, and 
$\ch^\pm_1\sim\ti W^\pm$ are kinematically allowed. 
Therefore, for $\cstau\sim 0$ $\stau_1$ decays predominately into 
$\tau\nt_1$ while for $|\cstau|\sim 1$ it mainly decays into  $\tau\nt_2$
and $\snu_\tau \ch^-_1$. $\stau_2$ shows the opposite behaviour. 
For the $\snu_\tau$ decays we obtain 
BR$(\snu_\tau\to \nu\nt_1)=32\%$, 
BR$(\snu_\tau\to \nu\nt_2)=17\%$, and 
BR$(\snu_\tau\to \tau\ch^+_1)=51\%$
for $m_{\snu_\tau}=148$ GeV and the other parameters as 
in Fig.~\ref{fig:Xsectsl}. 
This means that at least 1/3 of the events are invisible.

\begin{figure}[t]
\graph{0cm}{0cm}{-0.6cm}{!}{!}{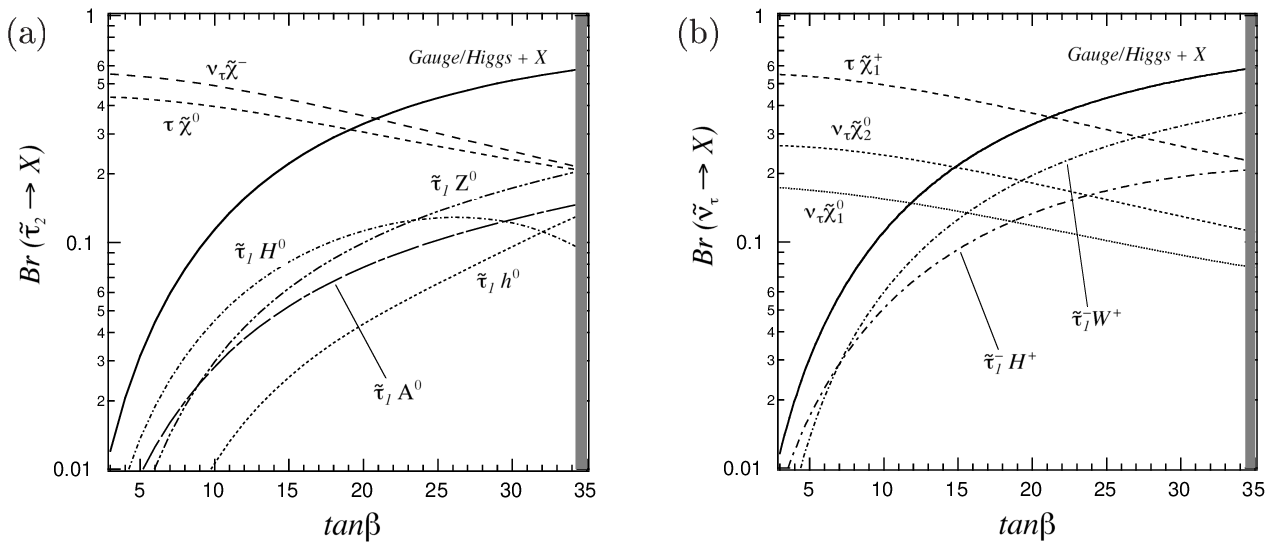}
\caption{$\tan\b$ dependence of $\stau_2$ (a) and $\snu_\tau$ (b) decay 
branching ratios for $\mstau{1}=250$ GeV, $\mstau{2}=500$ GeV, 
$\mstau{L}<\mstau{R}$, $A_\tau=800$ GeV and $\mu=1000$ GeV; 
the other parameters are:
$M=300$ GeV,  $m_A=150$ GeV, and 
$M_{\ti Q}=M_{\ti U}=M_{\ti D}=A_t=A_b=500$ GeV.
\label{fig:BRstau2}}
\end{figure}

\noi
In case of a large mass splitting $\mstau{2}-\mstau{1}$, $\stau_i$ and 
$\snu_\tau$ can also decay into gauge or Higgs bosons. 
This is especially the case if $\tan\b$, $A_\tau$ and $\mu$ are large. 
As an example, 
Fig.~\ref{fig:BRstau2} shows the branching ratios of $\stau_2$ and $\snu_\tau$ 
decays as a function of $\tan\b$ for $\mstau{1}=250$ GeV, $\mstau{2}=500$ GeV, 
$\mstau{L}<\mstau{R}$, $A_\tau=800$ GeV and $\mu=1000$ GeV, 
$M=300$ GeV, and $m_A=150$ GeV. 
(``Gauge/Higgs + X'' refers to the sum of the gauge and Higgs boson modes.)
As can be seen, with increasing $\tan\b$ the bosonic decay modes become dominant. 
See \cite{slbosdec} for more details.

\section{Parameter determination}

\begin{figure}[t]
\graph{0cm}{0cm}{-0.4cm}{!}{!}{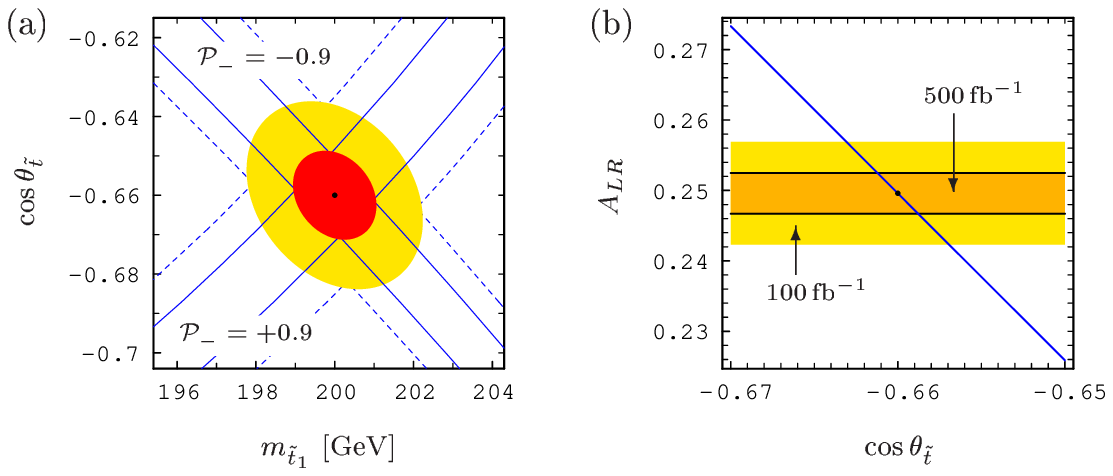}
\caption{(a) Error bands and 68\% CL error ellipse for determining $\mst{1}$ 
and $\cst$ from cross section measurements;  
the dashed lines are for ${\cal L}=100\;\fbi$ and 
the full lines for ${\cal L}=500\;\fbi$. 
(b) Error bands for the determination of $\cst$ from $A_{LR}$.
In both plots $\mst{1}=200$ GeV, $\cst=-0.66$, $\sqrt{s}=500$ GeV, 
$\Pm=\pm 0.9$, $\Pp=0$, 
and the other parameters as in Fig.~\ref{fig:Xsectsq}.
\label{fig:pol-err}}
\end{figure}

We next estimate the precision one may obtain for the parameters 
of the $\st$ sector from cross section measurements. 
We use the parameter point of Fig.~\ref{fig:Xsectsq},
i.e. $\mst{1}=200$ GeV, $\mst{2}=420$ GeV, 
$\cst=-0.66$, etc. as an illustrative example: 
For 90\% left--polarized electrons (and unpolarized positrons) we have  
$\s_L(\st_1\bar\st_1)=44.88\;$fb, 
including SUSY--QCD, Yukawa coupling, and ISR corrections.  
For 90\% right--polarized electrons we have $\s_R(\st_1\bar\st_1)=26.95\;$fb.
%
Assuming that $M$, $\mu$, $\tan\b$, and $m_A$ will be known from other 
measurements within a precision of about 10\% and taking into account 
$\d{\cal P}/{\cal P}\simeq 10^{-2}$ leads to an uncertanity of  
these cross sections of $\Delta\s/\s \lsim 1\%$. 
(Higher order QCD effects may add to this uncertanity; however, they 
have not yet been calculated.)
According to the Monte Carlo study of \cite{nowak} 
one can expect to measure the $\st_1\bar\st_1$ production cross sections 
with a statistical error of $\D\s_L/\s_L = 2.1\,\%$ and $\D\s_R/\s_R = 2.8\,\%$ 
in case of an integrated luminosity of ${\cal L}=500\;\fbi$ 
(i.e. ${\cal L}=250\;\fbi$ for each polarization).
Scaling these values to ${\cal L}=100\;\fbi$ leads to 
$\D\s_L/\s_L = 4.7\,\%$ and $\D\s_R/\s_R = 6.3\,\%$.  
Figure~\ref{fig:pol-err}\,a shows the corresponding error bands and 
error ellipses in the $\mst{1}$--$\,\cst$ plane. 
The resulting errors on the stop mass and mixing angle are:  
$\D\mst{1}=2.2$ GeV, $\D\cst= 0.02$ for ${\cal L}=100\;\fbi$ 
and $\D\mst{1}=1.1$ GeV, $\D\cst= 0.01$ for ${\cal L}=500\;\fbi$. 
With the additional use of a 60\% polarized $e^+$ beam these values can still 
be improved by $\sim 25\%$. 
At $\sqrt{s}=800$ GeV also $\st_2$ can be produced: 
$\s(\st_1\bar\st_2+\mbox{c.c.})=8.75$\,fb for $\Pm=-0.9$ and $\Pp=0$. 
If this cross section can be measured with a precision of $6\,\%$   
this leads to $\mst{2}=420\pm 8.9$ GeV 
(again, we took into account a theoretical uncertainity of 1\%).\footnote{Here 
  note that $\st_1\bar\st_1$ is produced at $\sqrt{s}=800$ GeV with 
  an even higher rate than at $\sqrt{s}=500$ GeV. 
  One can thus improve the errors on $\mst{1}$, $\mst{2}$, and $\cst$ 
  by combining the information obtained at different energies. 
  However, this is beyond the scope of this study.} 
With $\tan\b$ and $\mu$ known from other measurements this then allows one to 
determine the soft SUSY breaking parameters of the stop sector.  
Assuming $\tan\b=4\pm 0.4$ leads to $M_{\ti Q}=298\pm 8$ GeV and 
$M_{\ti U}=264\pm 7$ GeV for ${\cal L}=500\,\fbi$.
In addition, assuming $\mu=800\pm 80$ GeV we get $A_t=587\pm 35$ 
(or $-187 \pm 35$) GeV.
The ambiguity in $A_t$ exists because the sign of 
$\cst$ can hardly be determined from cross section measurements. 
This may, however, be possible from measuring decay branching ratios 
or the stop--Higgs couplings. 

A different method to determine the sfermion mass is to use kinematical 
distributions. 
This was studied in \cite{feng} for squarks of the 
1st and 2nd generation. It was shown that, by fitting the distribution 
of the minimum kinematically allowed squark mass, 
it is possible to determine $m_{\sq}$ with high precision. 
To be precise, \cite{feng} concluded that at $\sqrt{s}=500$ GeV,  
$m_{\sq}\sim 200$ GeV could be determined with an error of $\lsim 0.5\%$ 
using just $20\,\fbi$ of data (assuming that all squarks decay via 
$\sq\to q\nt_1$ and $\mnt{1}$ is known). 
The influence of radiative effects on this method has been studied 
in \cite{degk}. 
Taking into account initial state radiation of photons and gluon radiation 
in the production and decay processes it turned out that a mass of 
$m_{\sq}=300$ GeV could be determined with an accuracy of $\lsim 1\%$ with 
$50\,\fbi$ of data. 
(This result will still be affected by the error on the assumed $\mnt{1}$, 
hadronization effects, and systematic errors.) 
Although the  analysis of \cite{feng,degk} was performed for squarks of the 1st
and 2nd generation, the method is also applicable to the 3rd generation. 

For the determination of the mixing angle, one can also make use of 
the left--right asymmetry $A_{LR}$, Eq.~(\ref{eq:alr}). 
This quantity is of special interest because kinematic effects and 
uncertainities in experimental efficiencies largely drop out. 
At $\sqrt{s}=500$~GeV we get $A_{LR}(e^+e^-\to \st_1\bar{\st_1})=0.2496$ 
for the parameter point of Fig.~\ref{fig:Xsectsq} 
and 90\% polarized electrons. 
Taking into account experimental errors as determined in \cite{nowak},  
a theoretical uncertanity of 1\%, and $\delta P/P=10^{-2}$ we 
get $\Delta A_{LR}=2.92\%$ (1.16\%) for ${\cal L}=100\,\fbi\;(500\,\fbi)$. 
This corresponds to $\Delta\cst=0.0031$ (0.0012). 
This is most likely the most precise method to determine the stop mixing 
angle. 
The corresponding error bands are shown in Fig.~\ref{fig:pol-err}\,b.

A Monte Carlo study of stau production, with $\stau_1\to\tau\nt_1$, 
was performed in \cite{nojiri}. They also give a method for 
the parameter determination concluding that $m_{\stau_1}$ and $\t_{\stau}$ 
could be measured with an accurracy of ${\cal O}(1\%)$.

\section{Comparison with LHC and Tevatron}

In this section we briefly discuss the possibilites of detecting 
(light) stops, sbottoms, and staus at the LHC or Tevatron. 
At hadron colliders, stops and sbottoms are produced in pairs via 
gluon--gluon fusion or $q\bar q$ annihilation. 
They are also produced singly in gluon--quark interactions. At leading order, 
the production cross sections depend only on the masses of the particles 
produced. The NLO corrections introduce a dependence on the other MSSM 
parameters of ${\cal O}(1\%)$ \cite{beenakker98}. 
In addition, stops and sbottoms can be produced in cascade decays e.g., 
$\sg\to t\st_i$, $\sg\to b\sb_j$ with $\sb_j\to \st_i W^-$, 
$\sq\to q\bar q\nt_j$ with $\nt_j\to t\st_i$ etc. 

\noi
At the LHC one is, in general, sensitive to squark masses up to 
$\sim 2$ TeV \cite{cms,atlas}.
Searches for stops, however, suffer from an overwhelming background from 
top quarks, which makes the analysis very difficult. Here notice that 
e.g., $\s(pp\to\st_1\bar{\st_1})\sim \frac{1}{10}\s(pp\to t\bar t)$ 
for $\mst{1}\sim m_t$ and  
$\s(pp\to\st_1\bar{\st_1})\sim \frac{1}{100}\s(pp\to t\bar t)$ 
for $\mst{1}\sim 300$ GeV. 
Therefore, \cite{gianotti} concluded that it is `extremely difficult' 
to extract a $\st$ signal if the SUSY parameters are [similar to] those 
of LHC Point 4, i.e.  
$m_0=800$ GeV, $m_{1/2}=200$ GeV, $A=0$, $\tan\b=10$, and $\mu>0$,  
leading to $\mst{1}=594$ GeV and $\msg=582$ GeV. 
The situation is more promising for LHC Point 5, i.e. 
$m_0=100$ GeV, $m_{1/2}=300$ GeV, $A=300$ GeV, $\tan\b=2.1$, and $\mu>0$,  
leading to $\mst{1}=490$ GeV and $\msg=770$ GeV. 
In this case $\mst{1}$ can be determined with an accurracy of 
$\sim 10\%$ \cite{polesello}. 
However, no information on $\tst$ is obtained. 
More importantly, in \cite{dydak} it turned out that a light $\st_1$ 
with $\mst{1}\lsim 250$ GeV is extremely difficult to observe 
at LHC.

\noi
Such a light $\st_1$ could, in principle, be within the reach of the 
Tevatron Run~II. 
A rather complete study of the Tevatron potential for stop searches 
was performed in \cite{demina}. It turned out that the reach in $\mst{1}$ 
depends very much on the decay channel(s) and kinematics and, of course, on 
the luminosity. 
For example, for $\st_1\to b\,\ch^+_1$ with $\cL=2\,\fbi$, it is not
possible to go beyond the LEP2 limit of $\mst{1}\gsim 100$ GeV. 
With $\cL=20\,\fbi$ the reach extends up to $\mst{1}=175$ (212) GeV 
if $\mch{1}=130$ (100) GeV. 
If $\st_1$ decays into $c\nt_1$, with $\cL=2\,\fbi$ one can exclude 
$\mst{1}\lsim 180$ GeV provided $\mnt{1}\sim 100$; 
with $\cL=20\,\fbi$ one can exclude $\mst{1}\lsim 225$ GeV 
if $\mnt{1}\lsim 135$ GeV.
Notice, however, that no limit on $\mst{1}$ can be obtained with 
$\cL=2$ (20) $\fbi$ if $\mnt{1}\gsim 110$ (140) GeV 
or if $\mst{1}-\mnt{1}\lsim 15$ GeV. 
If $\mnt{1}\lsim \mst{1}-m_b-m_W$ the decay $\st_1\to bW\nt_1$ becomes 
relevant and one can hardly exceed the limits from LEP searches, 
even not with $\cL=20\,\fbi$. 
Similar results have been obtained for $\st_1$ three--body decays into 
sleptons. 


\noi 
The authors of \cite{demina} also studied the search for light sbottoms 
at the Tevatron Run~II concentrating on the decay $\sb_1\to b\nt_1$ 
within mSUGRA. 
They conclude that with $2\,\fbi$ of data the reach is 
$\msb{1}\lsim 200$ (155) GeV for $\mnt{1}\simeq 70$ (100) GeV. 
With $20\,\fbi$ one is sensitive to $\msb{1}\lsim 260$ (200) GeV for 
$\mnt{1}\simeq 70$ (100) GeV. 
Moreover, their analysis requires a mass difference of 
$\msb{1}-\mnt{1}\gsim 30$ GeV.
The higher reach compared to $\st_1\to c\nt_1$ is due to the 
higher tagging efficiency of $b$'s. 
Similarly, also at the LHC the search for sbottoms is, in general, 
expected to be easier than that for stops. There are, however, cases 
where the analysis is very difficult, see e.g. \cite{gianotti}.


The search for staus crucially depends on the 
possibility of $\tau$ identification. 
At hadron colliders, $\stau$'s are produced directly 
via the Drell--Yan process mediated by $\gamma$, $Z$ or $W$ exchange in 
the $s$--channel. They can also be produced in decays of charginos 
or neutralinos originating from squark and gluino cascade decays e.g., 
$\sq\to q'\ch^\pm_j$ with $\ch^\pm_j\to\stau^\pm_i \nu_\tau$ or 
$\ch^\pm_j\to\snu_\tau \tau^\pm$. 
At Tevatron energies, $W$ pair production is the dominant background, 
while $t\bar t$ events, with the $b$ jets being too soft to be detected, 
are the main background at the LHC. SUSY background mainly comes from 
$\ch^\pm\ch^\mp$ production followed by leptonic decays. 
The Drell--Yan production has a low cross section, and it is practically 
impossible to extract the signal from the SM background 
(SUSY background is less important). 
The situation is different if chargino and neutralino decays into 
staus have a large branching ratio. 
As pointed out in \cite{baer97,baer98,lali-paper} this is the case for 
large $\tan\b$ where the tau~Yukawa coupling becomes important. 
In \cite{baer98,baer99,atlas} the decays 
$\ch^\pm_1\to\stau_1^{}\nu_\tau$ and $\nt_2\to\stau_1^{}\tau$
($\stau_1^{}\to \tau\nt_1$) with the $\tau$'s decaying hadronically 
have been studied. 
In \cite{lali-paper,lali-diss}, the dilepton mass spectrum of 
final states with $e^+e^-/\mu^+\mu^-/e^\pm\mu^\mp + E_T^{miss} + jets$ 
has been used to identify $\stau_1$ in the decay chain
$\nt_2\to\stau_1^{}\tau\to\nt_1\tau^+\tau^-$ with 
$\tau\to e\,(\mu)+\nu_{e\,(\mu)}+\nu_\tau$. 
It turned out that $\stau_1^{}$ with $\mstau{1}\lsim 350$ GeV 
ought to be discovered at the LHC if  
$\mstau{1} < \mnt{2}$ and $\tan\b\gsim 10$. 

\noi
From this one can conclude that there exist MSSM parameter regions for 
which (light) sfermions of the 3rd generation may escape detection at 
both the Tevatron and the LHC. 
This is in particular the case for $\st_1$ if $\mst{1}\lsim 250$ GeV,  
and for $\stau_1^{}$ if $\tan\b\lsim 10$ (or $\mstau{1} > \mnt{2}$).
In these cases, an $e^+e^-$ Linear Collider 
would not only allow for precision measurements 
but even serve as a discovery machine.

\section{Summary}
In this contribution we discussed the phenomenology of stops, sbottoms, 
$\tau$--sneutrinos, and staus at an $e^+e^-$ Linear Collider with 
$\sqrt{s} = 0.5-1$~TeV.
We presented numerical predictions within the Minimal 
Supersymmetric Standard Model for the production cross sections 
and the decay rates of these particles, and analyzed their SUSY parameter 
dependence. 
Beam polarization turned out to be a very useful tool: 
Firstly, the dependence of the production cross sections  
on the sfermion mixing angles 
is significantly stronger if polarized beams are used. 
Secondly, one could enhance the production of $\sf_2$ pairs and 
reduce at the same time the production of $\sf_1$ pairs or vice versa. 
In such a case a better separation of the two mass eigenstates is
possible.
Concerning the decays, we showed that squarks and sleptons of the 3rd 
generation can have quite complex decay patterns. In particular, we 
discussed higher--order decays of $\st_1$. Moreover, we showed that 
for $\st$, $\sb$, $\stau$ and $\snu_\tau$ decays into lighter sfermions 
plus gauge or Higgs bosons can have large branching ratios. 
We also made a case study for the determination of the MSSM parameters of 
the $\st$ sector, showing that a precision of few percent may be achieved 
at the Linear Collider.    
Comparing with LHC and Tevatron, a light $\st_1$ ($\mst{1}\lsim 250$ GeV) 
may escape detection at the hadron colliders. 
In this case it will be discovered at a Linear Collider 
with $\sqrt{s}=500$ GeV.  
Also the detection of $\stau$'s is possible at the LHC only in a quite  
limited parameter range whereas it should be no problem at the Linear 
Collider.

\section*{Acknowledgements}

We thank the organizers of the various ECFA/DESY meetings for 
creating an intimate and inspiring working atmosphere. 
We also thank H.U. Martyn and L. Rurua for fruitful discussions and 
suggestions. 
This work was supported in part by the ``Fonds zur F\"orderung der 
Wissenschaftlichen Forschung of Austria'', project no. P13139-PHY. 
W.P.~is supported by the Spanish ``Ministerio de Educacion y Cultura'' 
under the contract SB97-BU0475382, by DGICYT grant PB98-0693, and by the 
TMR contract ERBFMRX-CT96-0090.


\end{document}